# Gender expression appraisals in introductory physics courses: A cross-institutional replication

Yangqiuting Li and Noah Leibnitz
*Department of Physics, Oregon State University, Corvallis, Oregon 97331, USA*

**Abstract**

Quantitative studies of gender in physics education have often used categorical gender identity measures, which have been valuable for documenting broad patterns of inequity across groups. However, categorical measures are less well suited for capturing variation within gender groups or for examining how students perceive themselves and how they believe others perceive them along dimensions commonly associated with gender, such as femininity, masculinity, and androgyny. Gradational self-appraisal and reflected-appraisal measures of these dimensions offer a complementary approach. Prior work using this approach in introductory physics identified substantial within-gender variation in students' self-appraisals and gender-patterned self–reflected appraisal discrepancies, including tendencies toward positive masculinity, negative femininity, and positive androgyny discrepancies. Building on this work, the present study provides a cross-institutional replication by examining continuities and variations in these quantitative patterns in a second institutional context. We examined students' self-appraisals and reflected appraisals of femininity, masculinity, and androgyny, self–reflected appraisal discrepancies, and associations between these discrepancy patterns, sense of belonging, and gender stigma consciousness. Across institutional contexts, both studies showed substantial within-gender variation in all three appraisal dimensions and recurring directional discrepancy patterns. In the present study, these directional patterns were more prevalent among students who identified as women or were in gender identity groups with smaller sample sizes than among men. Higher gender stigma consciousness was consistently associated with all three directional discrepancy patterns across institutions. Lower sense of belonging was consistently associated with negative femininity discrepancy across institutions and was also associated with positive masculinity discrepancy in the present study. At the same time, some details differed across contexts, including discrepancy sizes, subgroup patterns, and specific predictor associations. These findings suggest that students' self-appraisals and reflected appraisals along gendered dimensions are both patterned and context-sensitive. More broadly, self–reflected appraisal discrepancy may offer a useful quantitative lens for examining perceived gaps between students' self-understandings and how they believe they are perceived by others. Because these discrepancies are associated with belonging and gender stigma consciousness in physics courses, this lens may help identify inclusion-relevant dynamics and inform efforts to broaden how students' expression, participation, and competence are recognized in physics learning environments.

## I. Introduction

Despite ongoing initiatives aimed at increasing participation, gender diversity in physics remains limited. For example, even though women earn approximately 60% of all bachelor's degrees in the U.S. [1], they earn only about 25% of physics bachelor's degrees [2]. Prior research has also documented gender disparities in performance [3–19], motivational beliefs [18,20–26], and other educational outcomes [27–29], motivating calls for instructional and structural reform. Much prior quantitative work has examined these disparities through comparisons across gender identity groups. While such comparisons have been critical for documenting broad patterns of inequity, they are less well suited for capturing variation within gender groups [30–32] or for examining how gender is enacted and interpreted in physics learning environments [33–35]. Prior PER scholarship has called for moving beyond binary models of gender and examining gender as complex, fluid, and socially situated within physics learning environments [31].

Gender expression provides a complementary lens for examining these aspects of students' experiences. Gender expression refers to how individuals communicate their gender to others, which may or may not align with societal expectations or norms [36]. In physics education research, this lens has been applied primarily in qualitative studies that examine students' lived experiences in physics. For example, prior work has shown that women in physics often feel pressure to distance themselves from stereotypically feminine traits in order to fit into predominantly masculine norms of expression [35,37,38]. Similarly, studies have documented that gender minority students report experiencing pressure to conform or to conceal aspects of their gender expression to avoid exclusionary experiences such as misgendering, harassment, or social isolation [39–42]. Collectively, these findings suggest that gender expression is an important dimension of students' experiences of inclusion and belonging in physics learning environments [35,37,40,41].

Building on this qualitative literature, our prior work examined whether students' appraisals along dimensions associated with gender expression could also be studied quantitatively in introductory physics courses [43]. To do so, we adapted gradational measures from prior sociological studies [30,32] to introductory physics courses, capturing students' self-appraisals of femininity, masculinity, and androgyny (how they appraised themselves on these dimensions) as well as reflected appraisals (how they believed others perceived them on these dimensions). Our study revealed substantial within-group variation in students' self-appraised femininity, masculinity, and androgyny, with responses spanning the full range of the measurement scale [43]. It also identified a novel pattern: many students reported discrepancies between their self-appraised and reflected appraisals of these dimensions. These appraisal discrepancies were not randomly distributed; certain directional discrepancies were more prevalent among women and students of color and were associated with gender stigma consciousness and, in some cases, sense of belonging in physics [43].

Because these findings emerged from a single institutional context, the scope of their applicability remained unclear [43]. In particular, it was not clear whether the observed patterns reflected broader features of physics learning environments or dynamics shaped by local cultural and institutional conditions. Recent methodological work in physics education highlights that replication is especially important in educational research because it helps clarify whether prior findings remain robust across contexts and study conditions, thereby contributing to cumulative knowledge [44]. In this study, we therefore conduct a replication at a second large public research university in a different region of the United States. Using the same core gradational measures and a closely aligned analytic approach, we examine whether key patterns identified in prior work recur in a new institutional context. By identifying which patterns replicate and which vary across institutions, this study contributes to a more robust empirical understanding of gender expression appraisals and appraisal discrepancies in physics learning environments.

## II. Background

### A. Gender inequities in physics

A substantial body of research has documented persistent gender-related inequities in physics, spanning participation, academic outcomes, and social-psychological experiences [45]. At the level of representation, women and other historically marginalized groups remain underrepresented across multiple stages of physics training and careers [2,45], reflecting historical and ongoing barriers to access and recognition. In undergraduate physics, prior studies have also reported disparities in performance on several widely used conceptual inventories [10,11,46], and in some contexts on exams and course grades [4,6,8]. This literature emphasizes that such patterns are shaped by unequal opportunities and by features of assessment and instruction, rather than simply reflecting individual ability differences [3,5,31].

In parallel, a large body of work has documented inequities in students' social-psychological experiences that are closely tied to persistence, such as physics self-efficacy, interest, perceived recognition, physics identity, and sense of belonging [24,47–49]. These outcomes are linked to structural and cultural features of physics learning environments, such as limited access to role models and mentoring [48,50], a culture of competition and exclusion [51,52], and gender stereotypes about physics ability that frame physics success as requiring innate brilliance and associate competence with masculinity [53,54].

Importantly, gender inequity in physics is not limited to women–men comparisons. For example, in broader STEM contexts, gender minority students often report lower belonging and less positive perceptions of learning environments [39,55,56]. In physics, Barthelemy et al. found that gender-nonconforming respondents reported higher rates of discomfort in their physics classes, departments, and workplaces, and were significantly more likely than other respondents to report experiencing or observing exclusionary behavior [41,42].

### B. Gender expression as a social lens for studying physics learning environments

In response to these documented inequities, physics education researchers have increasingly sought to understand how students experience and navigate physics learning environments. In this context, gender expression provides an important lens for examining how gender is socially performed and interpreted within these environments [33,35,57]. Gender expression refers to how individuals present their gender through various means, such as behaviors, clothing, voice, and other characteristics [36]. Gender expression is not necessarily tied to a fixed identity; rather, it reflects how one's gendered sense of self is embodied and communicated to others [33,58]. In addition, gender expression is also dynamic, evolving culturally and over time [59,60].

In physics, several studies have examined the role of gender expression in students' experiences within the discipline. For example, Danielsson [35] investigated the experiences of female undergraduate students in Sweden, focusing on how these women navigate their gender identity and professional identity as physicists. The study suggests that many female students feel pressured to downplay their femininity to align with the masculine norms prevalent in physics. Similarly, Gonsalves [37] investigated how female doctoral students in physics navigate the tension between gender and competence. The study reveals that women in the field may struggle for recognition as competent physicists, often encountering scrutiny regarding their gender expression. This pressure can lead them to suppress aspects of their femininity to gain acceptance within the academic community.

In addition, prior research has also documented challenges related to gender expression among gender minority or gender-nonconforming individuals in physics. For example, studies situated within LGBTQ+ populations have shown that transgender and gender-nonconforming physicists often experience exclusionary behaviors tied to their gender expression, including misgendering, harassment, and social isolation[40–42]. These experiences frequently lead individuals to monitor, modify, or conceal aspects of their gender expression in order to reduce risk and maintain access to physics spaces [40,56,61].

### C. Dimensions of Gender Expression

Early research into gender expression primarily focused on two dimensions: femininity and masculinity [62]. Masculinity and femininity typically refer to traits, behaviors, and roles traditionally associated with men and women, respectively. Prior to 1974, most gender-related self-concept measures were constructed under the assumption that masculinity and femininity were two poles of the same trait, where an increase in one led to a decrease in the other [63]. However, this binary view was later challenged by researchers who proposed that femininity and masculinity are independent dimensions and individuals can express femininity and masculinity simultaneously [64]. This shift paved the way for the concept of androgyny, which refers to the possession of both masculine and feminine characteristics [64]. Importantly, the characteristics associated with gender expression can differ significantly across cultures. For instance, in some East Asian cultures, modesty and self-restraint—traits often seen as feminine in the West—are considered as masculine virtues, particularly in contexts that emphasize social harmony and collective well-being [65].

### D. Measuring gender expression appraisals with gradational self-appraisal and reflected-appraisal items

One of the most well-known instruments related to gender expression is the Bem Sex Role Inventory (BSRI) [64]. The original BSRI consisted of a 60-item index assigning feminine, masculine, and androgynous scores based on gendered trait ratings [64], and it was later abbreviated to a 30-item version. Importantly, these scores are derived from respondents' identification with traits that were culturally coded as feminine or masculine [64]. For example, identifying with traits such as gentleness or compassion increases one's femininity score, whereas identifying with assertiveness increases one's masculinity score. As a result, BSRI scores may reflect respondents' alignment with gendered trait stereotypes, rather than directly capturing their own gendered sense of self.

Another widely used approach involves self-reported appraisal measures, which allow individuals to directly rate their own levels of femininity, masculinity, and androgyny [32,66]. Unlike the BSRI, these measures allow respondents the flexibility to define how various characteristics relate to different dimensions of gender expression, rather than confining them to predetermined trait categories. One widely adopted self-reported measure is the gradational scale proposed by Magliozzi et al., which allows individuals to assess their levels of femininity and masculinity on two separate 7-point Likert scales [32]. Additionally, Magliozzi et al. proposed a measure of reflected appraisal, which assesses how individuals believe their femininity and masculinity are perceived by others [32]. In a later study, androgyny was added as another dimension [67], further broadening the scope of this measure. The measures developed by Magliozzi et al. have been adapted to investigate a range of topics, including gender experiences [68], perceived health [30], political attitudes [69], and economic decisions [70].

### E. Prior quantitative work and the need for cross-institutional replication

In our prior work, we adapted Magliozzi et al.'s gradational measures to introductory physics courses to examine students' self-appraisals and reflected appraisals of femininity, masculinity, and androgyny within physics learning environments [43]. We found that students' ratings of femininity, masculinity, and androgyny spanned the full range of the measurement scale, with substantial variation both across and within gender identity groups.

In addition to documenting variation in gender expression appraisals, the prior study showed that for many students, their appraisals of femininity, masculinity, and androgyny did not fully align with how they believed others perceived them [43]. These discrepancies also showed directional patterns across demographic groups. In calculus-based courses, where women were underrepresented, women tended to report perceiving themselves as more masculine and less feminine than they believed others perceived them. Across both algebra-based and calculus-based courses, women also tended to report perceiving themselves as more androgynous than they believed others perceived them. Similar patterns were observed across racial/ethnic groups: students of color were more likely than White students to report perceiving themselves as more masculine and more androgynous than they believed others perceived them.

Importantly, these appraisal discrepancies were associated with students' gender stigma consciousness and sense of belonging in physics [43]. Sense of belonging in physics refers to the extent to which students perceive themselves as valued, accepted, and legitimate members of physics [25], and prior research has linked belonging to students' academic performance, self-efficacy, motivation, and persistence [18,48,71–73]. Gender stigma consciousness refers to individuals' awareness of gender-based stigma and negative stereotypes, and has been linked to experiences of stereotype threat and academic disengagement [74]. In the prior study, students who perceived themselves as more masculine, more androgynous, or less feminine than they believed others perceived them tended to report higher gender stigma consciousness. Students who perceived themselves as less feminine than they believed others perceived them also tended to report lower sense of belonging [43]. These associations suggest that appraisal discrepancies may be meaningfully connected to students' experiences of inclusion in physics learning environments.

However, because this body of quantitative evidence was generated within a single institutional context [43], it remains unclear whether the observed patterns reflect broader features of physics learning environments or dynamics shaped by local cultural and institutional conditions. This question is especially relevant because gender expression is socially situated : how students understand and express their femininity, masculinity, and androgyny, and how these expressions are interpreted by others, may depend on societal expectations, disciplinary cultures, instructional practices, and institutional contexts [33,35,37,66,75].

In education research, replication is particularly useful for examining the scope of empirical findings across contexts [76,77]. Because educational phenomena are often context-sensitive, replication allows researchers to investigate whether observed patterns recur across settings or vary with institutional, cultural, or instructional conditions [76]. In this sense, replication can clarify the conditions under which prior findings hold across varied educational contexts. This is especially important when prior studies introduce new constructs or analytic approaches that may shape subsequent research and interpretation [76].

Therefore, the present study is a cross-institutional replication of prior quantitative work on students' self-appraisals and reflected appraisals of femininity, masculinity, and androgyny in introductory physics courses [43]. Using the same gradational measures and closely aligned analytic strategies, this study examines whether three key patterns recur in a new institutional context: (i) substantial within-group variation in self-appraisals of these dimensions, (ii) directional self–reflected appraisal discrepancy patterns, and (iii) relationships between discrepancy patterns and students' sense of belonging and gender stigma consciousness. By analyzing both cross-institutional similarities and differences, this study aims to clarify which aspects of these patterns recur across introductory physics environments and which appear sensitive to institutional context, thereby informing the interpretation and scope of findings about gender expression appraisals and appraisal discrepancy in introductory physics.

## III. Theoretical framework

The theoretical framework for this study follows the prior quantitative study and is grounded in Judith Butler's theory of gender performativity [34,57]. This theory challenges the understanding of gender as a fixed or inherent characteristic, instead conceptualizing gender as a sociocultural construct constituted and maintained through actions, expressions, and social interactions. From this perspective, gender is not something one simply "is," but something that is repeatedly performed and renegotiated within social contexts. Gender performativity further emphasizes that the meanings attached to gender are shaped by social norms, expectations, and institutional contexts rather than anchored in a stable binary system.

From a performativity perspective, individuals' understandings and expressions of gender are sensitive to the environments and contexts in which they are situated. Prior research has shown that different forms of work or social roles can shape how individuals perceive their own femininity or masculinity. For example, participation in care work has been associated with increased self-perceptions of femininity, whereas engagement in farm work has been associated with increased self-perceptions of masculinity [78,79]. More broadly, societal norms and ideals shape how individuals interpret and evaluate their own gender expression [80]. These findings underscore the importance of examining gender expression within specific social contexts.

Gender performativity theory also emphasizes the interactional nature of gender expression, which is not only communicated through individuals' actions but also perceived and interpreted by others within social interactions [81]. Prior research has shown that individuals' self-perceptions of gender do not always align with how others perceive them [82], and such misalignment has been associated with negative health outcomes among both cisgender and transgender people [30,82].

Consistent with this framework and with the prior study, we use gradational measures to examine students' self-appraisals and reflected appraisals of femininity, masculinity, and androgyny [32,43]. We also examine patterns of self–reflected appraisal discrepancy and how these discrepancies relate to students' broader social-psychological experiences in physics, including sense of belonging and gender stigma consciousness. Because gender expression is contextually situated, we compare the present results with those of the prior study to identify which patterns recur and which vary across institutional contexts.

## IV. Research questions

The present study is a replication of a prior quantitative study of gender expression appraisals in introductory physics [43]. The prior study was conducted at a large public research university in the southeastern United States, whereas the present study was conducted at a large public research university in the Pacific Northwest. Both studies included students from introductory algebra-based and calculus-based physics courses and used the same general measurement approach, including gradational self-appraisal and reflected-appraisal measures of femininity, masculinity, and androgyny, as well as social-psychological measures of sense of belonging and gender stigma consciousness. Our research questions are as follows.

**RQ1.** How do students with different gender identities in introductory physics courses report their self-appraisals of femininity, masculinity, and androgyny on gradational scales?

**RQ2.** Are there discrepancies between students' self-appraisals and reflected appraisals of femininity, masculinity, and androgyny? If so, how are these discrepancies associated with students' sense of belonging and gender stigma consciousness?

**RQ3.** To what extent do patterns in self-appraisals, self–reflected appraisal discrepancies, and associations between discrepancy patterns and sense of belonging and gender stigma consciousness recur or vary across institutional contexts?

## V. Methodology

### A. Participants and Course Context

The data used in this study were collected from students in introductory physics courses at a large public research university in the United States during one academic year. The institution follows a quarter system and offers both algebra-based and calculus-based introductory physics sequences. The algebra-based courses are typically taken by students majoring in life sciences, biomedical sciences, and other pre-health programs, whereas the calculus-based courses are typically taken by students majoring in engineering and the physical sciences. All courses included in this study were taught in person.

Both the algebra-based and calculus-based sequences consist of three introductory physics courses, referred to as Physics 1, Physics 2, and Physics 3, which cover comparable core physics topics. Physics 1 focuses on kinematics, mechanics, momentum, and energy; Physics 2 focuses on rotational motion, thermodynamics, fluids, oscillations, waves, and optics; and Physics 3 focuses on electricity and magnetism. While the two sequences differ in the level of mathematical formalism emphasized, the physics content and overarching learning goals are closely aligned.

The algebra-based introductory physics courses are five-credit courses and typically include approximately 3 hours of lecture per week, a 1-hour recitation, and a 2-hour laboratory. The calculus-based courses are four-credit courses and use two instructional formats. Some sections include approximately 2 hours of lecture per week, 2 hours of studio-style instruction, and a 2-hour laboratory, whereas other sections include approximately 3 hours of lecture per week and a 2-hour laboratory without studios. Lecture sections in both the algebra-based and calculus-based courses typically enroll approximately 100–200 students. Laboratory sections enroll approximately 30 students, with students typically working in small groups of two to three.

### B. Data collection

Survey data were collected from students enrolled in Physics 1, Physics 2, and Physics 3 in both the algebra-based and calculus-based introductory physics sequences during one academic year, spanning three consecutive academic terms. Surveys were administered during the last week of each course. Participation was voluntary, and students provided informed consent for their responses to be used for research purposes. Students who did not provide consent were excluded from the analytic sample. All study procedures were approved by the university's Institutional Review Board.

Across the three terms studied, approximately 550 students were enrolled in the algebra-based introductory physics courses and approximately 1990 students were enrolled in the calculus-based introductory physics courses. The final analytic sample included 1845 students, with 460 students from algebra-based courses and 1385 students from calculus-based courses. Not all enrolled students were included in the analytic sample because some did not participate in the survey, some participated but did not provide consent for their responses to be used for research purposes, and some students who were initially enrolled were no longer enrolled when the end-of-course survey was administered. Because the introductory physics curriculum is structured as a multi-course sequence, some students appeared in the dataset more than once if they were enrolled in multiple courses during the data collection period. For students with multiple survey responses, we retained only their final available response for the analyses reported in this study.

### C. Student demographics

Students' demographic information was obtained from survey responses. Gender identity was collected through self-report on the survey, which included multiple response options as well as an additional open-text option, "Gender not listed. My gender is:", through which respondents could describe their gender. In the algebra-based courses, most respondents identified as women (67.8%) or men (27.9%). Smaller percentages identified as non-binary (1.5%) or agender/I do not identify with any gender (1.2%), and 1.7% preferred not to state their gender. No respondents in the algebra-based courses selected the open-text gender option. In the calculus-based courses, most respondents identified as men (67.2%) or women (27.0%). Smaller percentages identified as non-binary (2.7%) or agender/I do not identify with any gender (0.3%), 2.4% preferred not to state their gender, and 0.5% selected the open-text gender option. Open-text gender responses in the calculus-based courses included terms such as "Genderfluid/nonbinary" and "Genderqueer."

Race and ethnicity were also collected through self-report on the survey, which included multiple response options as well as an additional option for respondents to select "Race/ethnicity not listed. My race/ethnicity is:" and provide their own written response. In the algebra-based courses, the most frequently selected race/ethnicity categories were White (62.4%), Asian (13.3%), Hispanic and Latino (10.8%), and Two or more races (7.1%). Smaller percentages identified as Black or African American (2.7%) or Native American or Alaska Native (0.6%), 2.1% preferred not to state, and 1.0% selected the open-text option. In the calculus-based courses, the most frequently selected race/ethnicity categories were White (64.0%), Asian (12.6%), Two or more races (8.8%), and Hispanic and Latino (7.6%). Smaller percentages identified as Black or African American (1.3%) or Native American or Alaska Native (0.9%), 2.9% preferred not to state, and 1.9% selected the open-text option. Across the full sample, the most common open-text responses to the race/ethnicity item included "Middle Eastern," "Arab," "Jewish," and "Pacific Islander."

### D. Measures

#### 1. Gradational measures of femininity, masculinity, and androgyny

In this study, we used gradational measures of femininity, masculinity, and androgyny adapted from prior work [30,32,67] to assess students' self-appraisals and reflected appraisals of these dimensions. The survey items and corresponding response scales are shown in Figure 1.

Students' self-appraisals were measured using the prompt *"In general, how do you see yourself?"*, which asked students to rate their femininity, masculinity, and androgyny on a 7-point Likert scale ranging from *not at all* (coded as 0) to *very* (coded as 6). Reflected appraisals were measured using a parallel prompt, *"In general, how do most people see you?"*, which asked students to rate how they believed others perceived their femininity, masculinity, and androgyny using the same response scale.

The self-appraisal and reflected-appraisal items were administered in the same survey as the social-psychological measures described below. The social-psychological measures appeared earlier in the survey, followed by the self-appraisal and reflected-appraisal items.

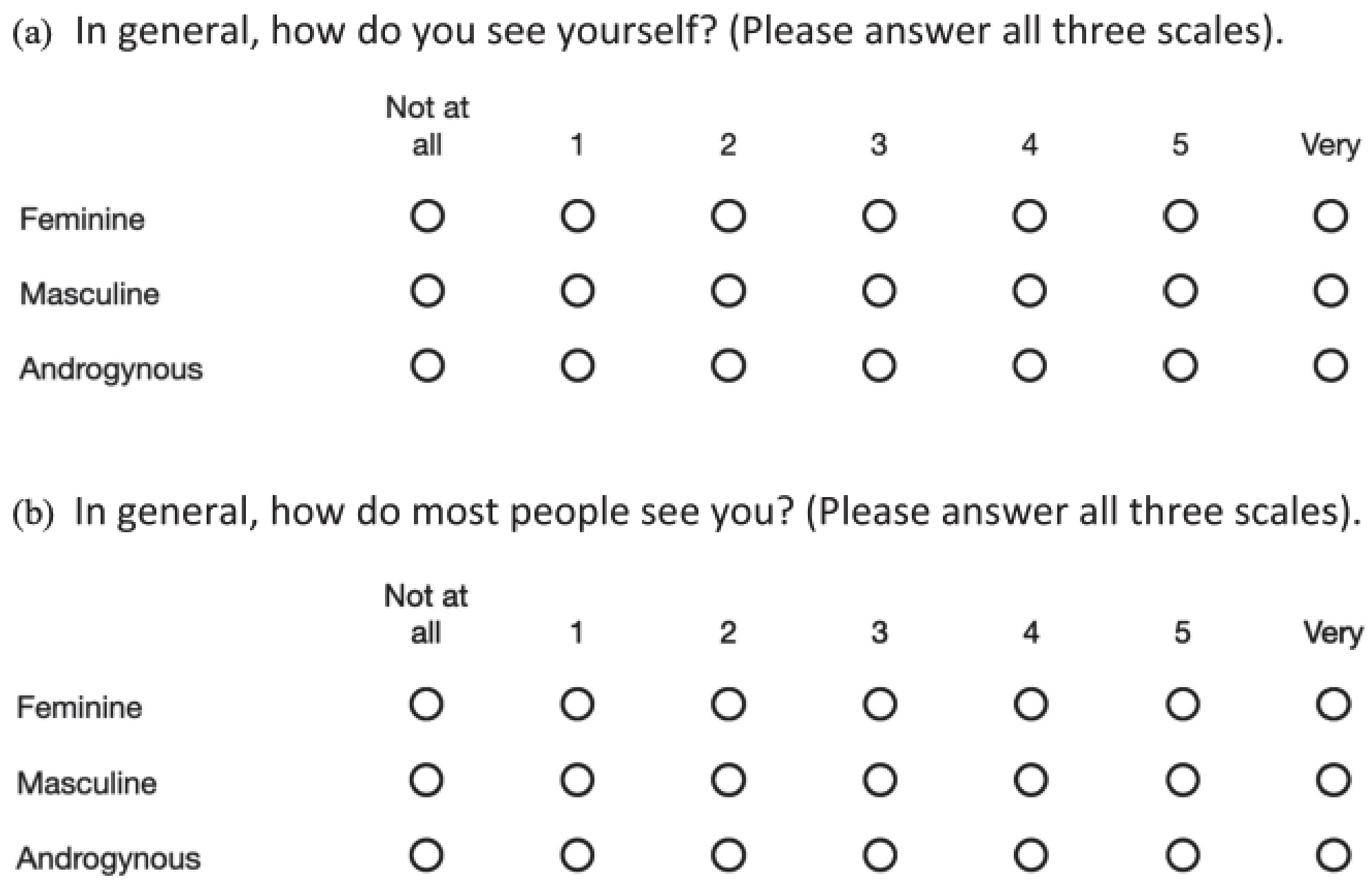

**Figure 1.** Survey items used to measure students' (a) self-appraisals and (b) reflected appraisals of femininity, masculinity, and androgyny. Responses were collected on a 7-point scale ranging from "Not at all" (0) to "Very" (6).

#### 2. Social-psychological measures

Following the prior study, we measured physics sense of belonging and gender stigma consciousness to examine associations between appraisal discrepancies and students' social-psychological experiences in physics. Both constructs were measured using items adapted from prior studies [18,83–86], with responses collected on a 7-point Likert scale. For each construct, responses to the relevant items were averaged to form a composite score, with higher values indicating greater sense of belonging or greater gender stigma consciousness. In the present sample, both scales demonstrated strong internal consistency (Cronbach's α = 0.87 for physics sense of belonging and α = 0.93 for gender stigma consciousness) [87]. These composite scores were used in subsequent analyses.

**TABLE I.** Survey items used to measure gender stigma consciousness and physics sense of belonging. Items marked with † were reverse-coded before analysis.

| **Construct and Item** |
| --- |
| **Gender Stigma Consciousness** (Cronbach's alpha = 0.93) |
| My gender affects how my peers interact with me. |
| Most people judge me on the basis of my gender. |
| **Physics Sense of Belonging** (Cronbach's alpha = 0.87) |
| I feel like I belong in this class. |
| I feel like an outsider in this class.† |
| I feel comfortable in this class. |
| I feel like I can be myself in this class. |
| Sometimes I worry that I do not belong in this physics class.† |

### E. Statistical analyses

We first examined the distributions of self-appraised femininity, masculinity, and androgyny using descriptive statistics and visualizations to characterize variability in students' appraisals of these dimensions. Distributions were examined separately for women and men. Students in gender identity categories with smaller sample sizes were combined for descriptive visualization because individual category sizes were too small to interpret separately. To complement these distributional analyses, we also examined the prevalence of endpoint responses (0 = *not at all*, 6 = *very*) on the femininity and masculinity scales among women and men, both overall and by course type. This endpoint analysis was limited to women and men because these groups supported more stable comparisons across course types. Differences across course types were evaluated using two-sample proportion z tests, with Cohen's $h$ reported as an effect size [88].

To examine whether students' self-appraisals aligned with their reflected appraisals, we computed discrepancy scores for femininity, masculinity, and androgyny by subtracting reflected appraisal from self-appraisal. Positive values indicate that students perceived themselves as higher on a given dimension than they believed others perceived them, whereas negative values indicate that students perceived themselves as lower on that dimension than they believed others perceived them. Because these discrepancy scores were discrete and included many tied values, we summarized them using binned categories and used Wilcoxon signed-rank tests [89] to assess whether discrepancies differed from zero within each gender-by-course subgroup.

We examined associations between discrepancy patterns and demographic variables, course type, and social-psychological variables, including sense of belonging and gender stigma consciousness. To do so, we defined three binary discrepancy indicators corresponding to the primary directional patterns identified in the prior study and examined in the present replication: positive masculinity discrepancy, negative femininity discrepancy, and positive androgyny discrepancy. We then calculated Pearson correlations between these binary discrepancy indicators and the demographic variables, course type, and social-psychological variables. To further examine how sense of belonging and gender stigma consciousness varied across discrepancy direction, we compared mean levels of these variables across negative, zero, and positive discrepancy groups for each dimension using one-way ANOVAs, followed by Tukey's HSD post hoc tests when omnibus effects were significant [90].

Finally, we estimated a series of regression models to examine how demographic variables, course type, and social-psychological variables jointly predicted discrepancy patterns. We first used multiple linear regression to examine whether demographic variables and course type predicted students' sense of belonging and gender stigma consciousness. Logistic regression models were then used to predict binary discrepancy indicators: positive masculinity discrepancy, negative femininity discrepancy, and positive androgyny discrepancy. Predictors were entered sequentially to assess whether adding course type and social-psychological variables improved model fit beyond models including only demographic variables.

To situate the present findings in relation to the prior study, we compared major patterns across the datasets from the two institutions. This comparison focused on three areas: distributions of gradational femininity, masculinity, and androgyny ratings; patterns of self–reflected appraisal discrepancies; and predictors of directional discrepancy indicators in logistic regression models.

## VI. Results

### A. Distribution of self-appraised femininity, masculinity, and androgyny

We begin by examining distributions of students' self-appraised femininity, masculinity, and androgyny. Figure 2 shows these distributions on the 7-point scale for students who identified as women or men. Across all three measures, respondents collectively used the full range of the scale. Women more frequently reported higher femininity and lower masculinity, whereas men more frequently reported higher masculinity and lower femininity. However, the distributions also showed overlap between women and men, indicating that students' self-appraised femininity and masculinity did not simply align with binary gender identity.

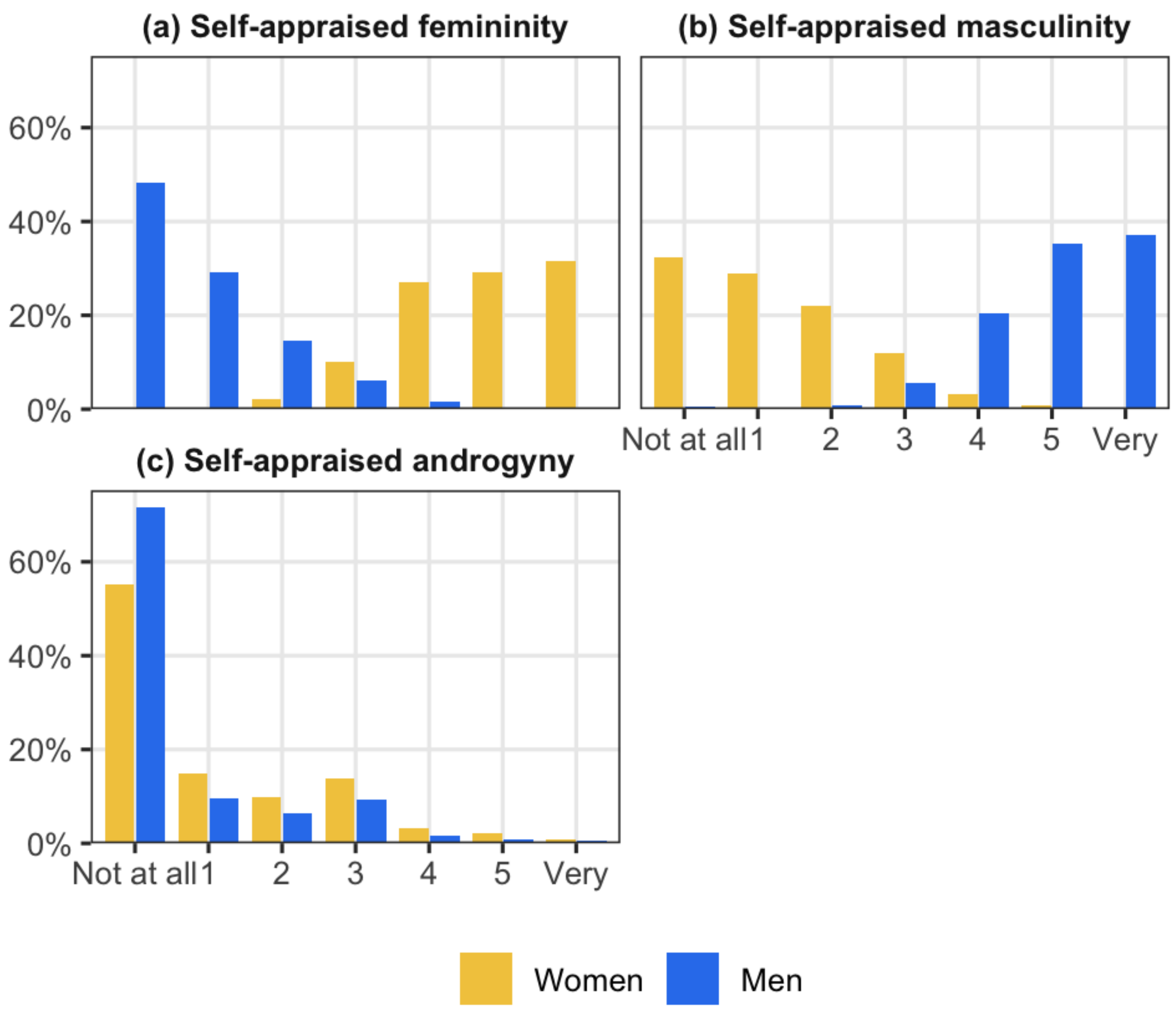

**Figure 2.** Distributions of self-appraised (a) femininity, (b) masculinity, and (c) androgyny for students who self-identified as women or men.

Figure 3 displays the corresponding distributions for students from gender identity groups with smaller sample sizes, including students who identified as non-binary, students who identified as agender, and students who provided self-described gender identities. These categories were combined for descriptive visualization because of their small sample sizes. Responses in this combined group were distributed primarily across intermediate levels of femininity and masculinity, with fewer responses at either end of these scales. In contrast, androgyny responses were more frequently reported at intermediate-to-high levels.

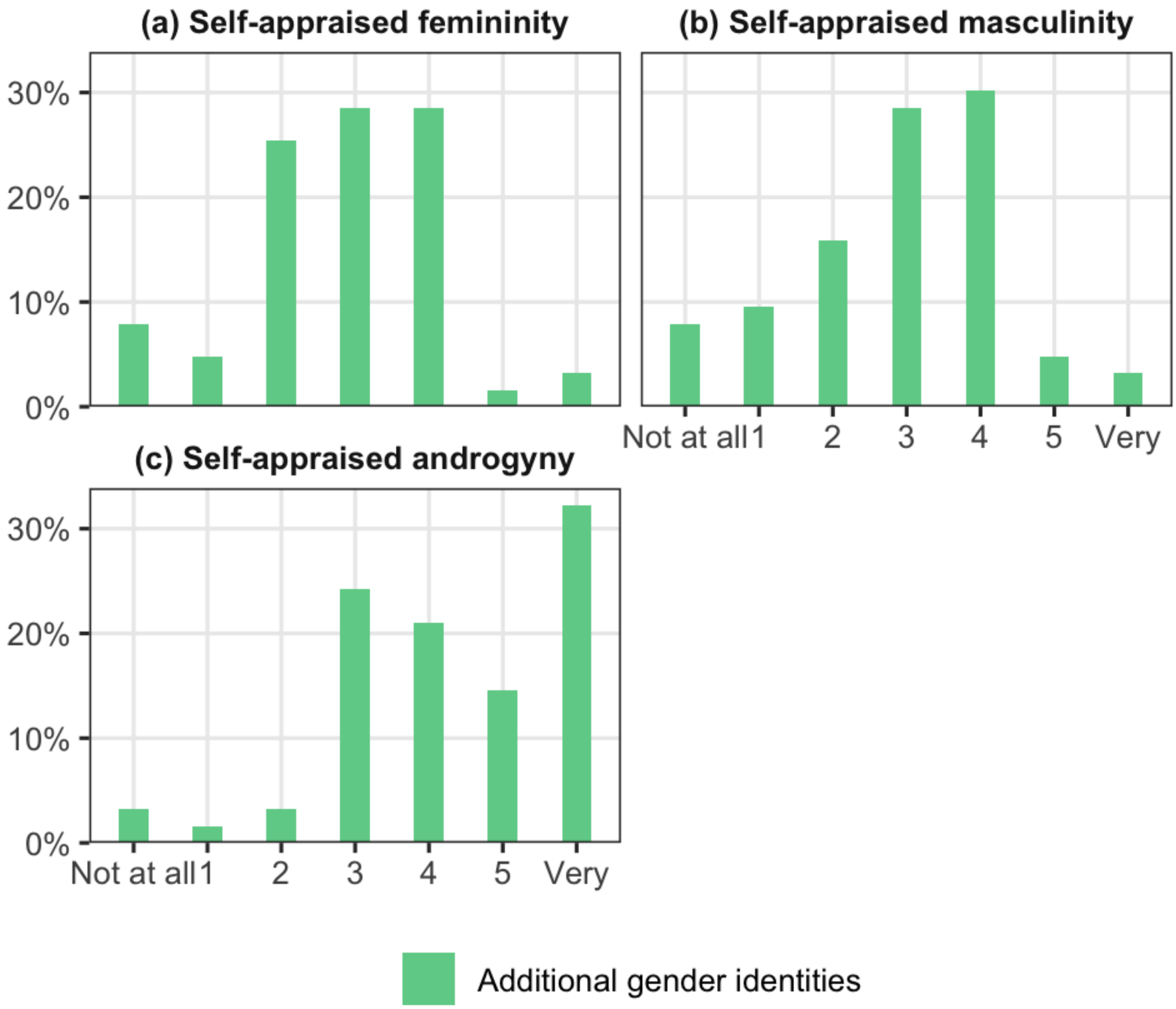

**Figure 3.** Distributions of self-appraised (a) femininity, (b) masculinity, and (c) androgyny for students in gender identity categories with smaller sample sizes, including non-binary, agender, and self-described gender responses.

To complement the distributional patterns shown in Fig. 2, Table II summarizes the percentages of men and women who selected the endpoints of the gradational femininity and masculinity scales (0 = not at all, 6 = very), both overall and by course type (algebra-based vs calculus-based). The table also reports results from two-sample proportion z tests comparing course types. Across both groups, most respondents did not select these endpoint values. For example, among women, 32% reported femininity = 6 and 32% reported masculinity = 0, with 22% selecting both femininity = 6 and masculinity = 0. We note that endpoint response patterns were largely consistent across course types; the only statistically significant difference was that a higher proportion of women in algebra-based courses than in calculus-based courses reported the simultaneous combination of femininity = 6 and masculinity = 0 ($p = 0.035$, $h = 0.16$). Collectively, the distributional and endpoint analyses show substantial variation in students' self-appraisals of femininity, masculinity, and androgyny both within and across gender identity groups.

**TABLE II.** Percentages of men and women who indicated their level of femininity and masculinity as 0 (not at all) or 6 (very), along with the results of two-sample proportion z tests to compare the percentages in the algebra-based and calculus-based courses. Cohen suggested that typically values of 0.2, 0.5, and 0.8 represent small, medium, and large effect sizes for two sample proportion z tests [88].

| Gender group | Endpoint response | Overall (N=1845) | Algebra-based (N=460) | Calculus-based (N=1385) | Statistics | |
|---|---|---|---|---|---|---|
| | | | | | *p-value* | Effect size (Cohen's *h*) |
| **Men** | Femininity = 0 | 48% | 50% | 48% | 0.766 | 0.03 |
| | Masculinity = 6 | 37% | 43% | 36% | 0.112 | 0.14 |
| | Femininity = 0 & Masculinity = 6 | 30% | 36% | 30% | 0.131 | 0.14 |
| **Women** | Femininity = 6 | 32% | 35% | 29% | 0.056 | 0.14 |
| | Masculinity = 0 | 32% | 36% | 30% | 0.068 | 0.14 |
| | Femininity = 6 & Masculinity = 0 | 22% | 26% | 19% | 0.035 | 0.16 |

## B. Comparison between students' self-appraisals and reflected appraisals of masculinity, femininity, and androgyny

Table III summarizes percentage distributions of differences between students' self-appraisals and reflected appraisals of masculinity in algebra-based and calculus-based courses, as well as overall (combining both course types). Positive values indicate higher self-appraisals than reflected appraisals, whereas negative values indicate lower self-appraisals than reflected appraisals; larger absolute values indicate larger discrepancies. The same discrepancy coding was used for femininity and androgyny in subsequent analyses.

Across both course types, women showed a tendency toward positive masculinity discrepancies. In calculus-based courses, 27% of women reported higher self-appraised than reflected masculinity, compared to 11% reporting the reverse. The Wilcoxon signed-rank test indicated a significant positive masculinity discrepancy ($p < 0.001$). A similar pattern was observed for women in algebra-based courses, with 26% reporting positive discrepancies and 9% reporting negative discrepancies ($p < 0.001$).

In contrast, men's self-appraisals and reflected appraisals of masculinity were largely aligned in both course types. The majority of men reported no discrepancy, and positive and negative discrepancies were approximately symmetric, yielding non-significant Wilcoxon signed-rank tests. Similarly, students from gender identity groups with smaller sample sizes, including non-binary, agender, and self-described gender identities, did not show statistically significant masculinity discrepancies in either course type.

**TABLE III.** Percentage distributions of differences between students' self-appraisals and reflected appraisals of masculinity in algebra-based and calculus-based courses, as well as overall, along with the results of Wilcoxon signed-rank tests assessing whether the median difference significantly differs from zero. For the r-type effect size reported for Wilcoxon signed-rank tests, values of 0.10, 0.30, and 0.50 are commonly interpreted as small, medium, and large effects, respectively [88]. The "smaller-sample gender groups" category includes students who identified as non-binary, agender, or self-described their gender; these categories were combined because of smaller sample sizes.

| | | Difference between self-appraisal and reflected appraisal of masculinity | | | | | | | | |
|---|---|---|---|---|---|---|---|---|---|---|
| | | **-3 and lower** | **-2** | **-1** | **0** | **+1** | **+2** | **+3 and higher** | *p-value* | *Effect size* |
| **Overall** | **Women** | 1% | 1% | 8% | 63% | 21% | 4% | 1% | <0.001 | 0.24 |
| | **Men** | 0% | 1% | 10% | 76% | 10% | 1% | 1% | 0.728 | 0.01 |
| | **Smaller-sample gender groups** | 6% | 11% | 13% | 24% | 27% | 10% | 10% | 0.379 | 0.08 |
| **Calculus-based** | **Women** | 1% | 2% | 8% | 62% | 23% | 3% | 1% | <0.001 | 0.21 |
| | **Men** | 0% | 1% | 10% | 76% | 9% | 1% | 1% | 0.665 | 0.01 |
| | **Smaller-sample gender groups** | 6% | 10% | 12% | 28% | 24% | 10% | 10% | 0.321 | 0.11 |
| **Algebra-based** | **Women** | 0% | 1% | 8% | 65% | 20% | 5% | 1% | <0.001 | 0.26 |
| | **Men** | 0% | 2% | 9% | 77% | 11% | 1% | 1% | 0.856 | 0.02 |
| | **Smaller-sample gender groups** | 8% | 15% | 15% | 8% | 38% | 8% | 8% | 0.936 | 0.02 |

Table IV summarizes percentage distributions of differences between students' self-appraisals and reflected appraisals of femininity in algebra-based and calculus-based courses, as well as overall. Overall, women showed a tendency toward negative femininity discrepancies, whereas men showed a tendency toward positive femininity discrepancies. Specifically, 20% of women reported negative discrepancies compared with 12% reporting positive discrepancies ($p < 0.001$, effect size = -0.13), whereas 16% of men reported positive discrepancies compared with 9% reporting negative discrepancies ($p = 0.002$, effect size = 0.10). Students from gender identity groups with smaller sample sizes also showed a negative femininity discrepancy pattern overall, with 56% reporting negative discrepancies and 21% reporting positive discrepancies (p = 0.006, effect size = -0.26).

Across course types, these patterns were not equally pronounced. Among women, negative femininity discrepancies were statistically significant in algebra-based courses (p = 0.001, effect size = -0.20) but not in calculus-based courses. Among men, positive femininity discrepancies were statistically significant in calculus-based courses (p = 0.006, effect size = 0.09) but not in algebra-based courses. Among students from gender identity groups with smaller sample sizes, a statistically significant negative femininity discrepancy was observed in calculus-based courses (p = 0.032, effect size = -0.23), whereas the corresponding pattern in algebra-based courses was not statistically significant.

**TABLE IV.** Percentage distributions of discrepancy scores between students' self-appraisals and reflected appraisals of femininity in algebra-based and calculus-based courses, as well as overall, along with results from Wilcoxon signed-rank tests.

| **Difference between self-appraisal and reflected appraisal of femininity** | | | | | | | | | | |
|---|---|---|---|---|---|---|---|---|---|---|
| | | **-3 and lower** | **-2** | **-1** | **0** | **+1** | **+2** | **+3 and higher** | ***p-value*** | ***Effect size*** |
| **Overall** | **Women** | 0% | 3% | 17% | 68% | 10% | 2% | 0% | <0.001 | -0.13 |
| | **Men** | 1% | 1% | 7% | 75% | 14% | 2% | 0% | 0.002 | 0.10 |
| | **Smaller-sample gender groups** | 13% | 19% | 24% | 24% | 10% | 5% | 6% | 0.006 | -0.26 |
| **Calculus -based** | **Women** | 0% | 3% | 18% | 64% | 12% | 2% | 1% | 0.082 | -0.09 |
| | **Men** | 1% | 1% | 7% | 75% | 13% | 2% | 0% | 0.006 | 0.09 |
| | **Smaller-sample gender groups** | 14% | 14% | 26% | 22% | 12% | 6% | 6% | 0.032 | -0.23 |
| **Algebra-based** | **Women** | 0% | 2% | 16% | 73% | 7% | 1% | 0% | 0.001 | -0.20 |
| | **Men** | 1% | 1% | 11% | 69% | 14% | 4% | 1% | 0.132 | 0.13 |
| | **Smaller-sample gender groups** | 8% | 38% | 15% | 31% | 0% | 0% | 8% | 0.080 | -0.38 |

Table V summarizes androgyny discrepancy patterns across gender groups and course types. Overall, positive androgyny discrepancies were observed across all gender groups, indicating that students tended to rate themselves as more androgynous than they believed others perceived them. This pattern was strongest among students from gender identity groups with smaller sample sizes (effect size = 0.34), followed by women (effect size = 0.20), and was smaller among men (effect size = 0.09).

Across course types, these patterns were more consistently observed in calculus-based courses. In calculus-based courses, positive androgyny discrepancies were statistically significant among women ($p < 0.001$, effect size = 0.19), men ($p = 0.004$, effect size = 0.09), and students in smaller-sample gender identity groups ($p < 0.001$, effect size = 0.49). In algebra-based courses, positive androgyny discrepancies remained statistically significant among women ($p < 0.001$, effect size = 0.24) and students in smaller-sample gender identity groups ($p = 0.010$, effect size = 0.59), but not among men.

Taken together, Tables III–V show three primary directional patterns: positive masculinity discrepancies among women, negative femininity discrepancies among women and students in smaller-sample gender identity groups, and positive androgyny discrepancies across gender groups, especially among women and students in smaller-sample gender identity groups.

**TABLE V.** Percentage distributions of discrepancy scores between students' self-appraisals and reflected appraisals of androgyny in algebra-based and calculus-based courses, as well as overall, along with results from Wilcoxon signed-rank tests.

| **Difference between self-appraisal and reflected appraisal of androgyny** | | | | | | | | | | |
|---|---|---|---|---|---|---|---|---|---|---|
| | | **-3 and lower** | **-2** | **-1** | **0** | **+1** | **+2** | **+3 and higher** | ***p-value*** | ***Effect size*** |
| **Overall** | **Women** | 1% | 1% | 5% | 73% | 14% | 4% | 2% | <0.001 | 0.20 |
| | **Men** | 1% | 1% | 4% | 84% | 8% | 2% | 1% | 0.004 | 0.09 |
| | **Smaller-sample gender groups** | 3% | 2% | 7% | 42% | 18% | 10% | 18% | <0.001 | 0.34 |
| **Calculus-based** | **Women** | 1% | 1% | 6% | 68% | 15% | 4% | 4% | <0.001 | 0.19 |
| | **Men** | 1% | 0% | 4% | 84% | 8% | 1% | 1% | 0.004 | 0.09 |
| | **Smaller-sample gender groups** | 2% | 0% | 2% | 31% | 33% | 6% | 27% | <0.001 | 0.49 |
| **Algebra-based** | **Women** | 0% | 1% | 4% | 78% | 14% | 2% | 0% | <0.001 | 0.24 |
| | **Men** | 2% | 1% | 5% | 81% | 9% | 3% | 0% | 0.536 | 0.05 |
| | **Smaller-sample gender groups** | 0% | 0% | 0% | 38% | 0% | 38% | 23% | 0.010 | 0.59 |

### C. Variables associated with self–reflected appraisal discrepancies in masculinity, femininity, and androgyny

The previous section showed that students' self-appraisals of femininity, masculinity, and androgyny did not always align with their reflected appraisals, and that these discrepancies exhibit directional patterns. In this section, we examine whether these discrepancy patterns are associated with demographic variables, course type, and social-psychological variables, including sense of belonging and gender stigma consciousness.

To examine these associations, we created binary indicators for the primary discrepancy directions observed in the preceding analyses. Each indicator was coded as 1 when the corresponding directional discrepancy was present and 0 otherwise: self-appraised masculinity > reflected masculinity, self-appraised femininity < reflected femininity, and self-appraised androgyny > reflected androgyny. Because students who identified as women and students in gender identity groups with smaller sample sizes showed broadly similar discrepancy patterns in the preceding analyses, and because both groups are historically underrepresented in physics [41,91], we combined them into a single binary gender variable for the purposes of the following association analyses. The variable Gender was coded 0 for students who identified as men and 1 for students who identified as women or were in gender identity groups with smaller sample sizes. Race/ethnicity was also represented as a binary variable, with students who identified as White coded as 0 and students who identified with other racial/ethnic groups coded as 1. Course type was coded as 0 for algebra-based courses and 1 for calculus-based courses.

Table VI presents correlations among the three discrepancy indicators and these potentially associated variables. The three discrepancy indicators were modestly and positively correlated with one another, suggesting that students who reported one type of discrepancy were somewhat more likely to report other directional discrepancy patterns. The Gender variable was positively correlated with positive masculinity discrepancy ($r = 0.21$, $p < 0.001$), negative femininity discrepancy ($r = 0.18$, $p < 0.001$), and positive androgyny discrepancy ($r = 0.18$, $p < 0.001$). These positive correlations indicate that these directional discrepancy patterns were more prevalent among students who identified as women or were in gender identity groups with smaller sample sizes than among men. Race/ethnicity and course type showed weaker associations with the discrepancy indicators.

Among the social-psychological variables, gender stigma consciousness showed the strongest and most consistent positive associations with the discrepancy indicators, with correlations ranging from 0.17 to 0.23 (all $p < 0.001$). Sense of belonging showed relatively small negative correlations with positive masculinity and negative femininity discrepancy indicators (both $r = -0.11$, $p < 0.001$) and little association with the positive androgyny discrepancy indicator.

**TABLE VI.** Pearson correlation coefficients among the three binary discrepancy indicators and the demographic, course-type, and social-psychological variables examined in the association analyses. The discrepancy indicators represent positive masculinity discrepancy, negative femininity discrepancy, and positive androgyny discrepancy. Statistical significance is indicated by *** for $p < 0.001$, ** for $0.001 \leq p < 0.01$, * for $0.01 \leq p < 0.05$, and ns for $p \geq 0.05$.

| Observed Variable | 1 | 2 | 3 | 4 | 5 | 6 | 7 | 8 |
|---|---|---|---|---|---|---|---|---|
| 1. Masculinity: self-appraisal > reflected appraisal | -- | -- | -- | -- | -- | -- | -- | -- |
| 2. Femininity: self-appraisal < reflected appraisal | $0.41^{***}$ | -- | -- | -- | -- | -- | -- | -- |
| 3. Androgyny: self-appraisal > reflected appraisal | $0.27^{***}$ | $0.18^{***}$ | -- | -- | -- | -- | -- | -- |
| 4. Gender | $0.21^{***}$ | $0.18^{***}$ | $0.18^{***}$ | -- | -- | -- | -- | -- |
| 5. Race/Ethnicity | $-0.02^{ns}$ | $0.01^{ns}$ | $-0.06^{*}$ | $0.02^{ns}$ | -- | -- | -- | -- |
| 6. Course type | $-0.06^{**}$ | $-0.05^{*}$ | $-0.01^{ns}$ | $-0.35^{***}$ | $-0.01^{ns}$ | -- | -- | -- |
| 7. Sense of belonging | $-0.11^{***}$ | $-0.11^{***}$ | $-0.02^{ns}$ | $-0.25^{***}$ | $-0.09^{***}$ | $0.13^{***}$ | -- | -- |
| 8. Gender stigma consciousness | $0.21^{***}$ | $0.17^{***}$ | $0.23^{***}$ | $0.35^{***}$ | $0.04^{ns}$ | $-0.02^{ns}$ | $-0.18^{***}$ | -- |

To further examine how students' social-psychological experiences varied across discrepancy directions, we compared mean sense of belonging and gender stigma consciousness across negative, zero, and positive discrepancy groups for masculinity, femininity, and androgyny. Figure 4 presents these comparisons for masculinity discrepancy.

Figure 4 shows that students in the positive masculinity discrepancy group—those whose self-appraised masculinity was greater than their reflected appraisal—reported the lowest sense of belonging and the highest gender stigma consciousness. A one-way ANOVA indicated significant group differences in sense of belonging, $F(2, 1854) = 12.20$, $p < 0.001$. Tukey post hoc tests showed that the positive discrepancy group reported significantly lower sense of belonging than both the zero discrepancy group ($p < 0.001$) and the negative discrepancy group ($p = 0.037$), whereas the negative and zero discrepancy groups did not differ significantly.

A parallel one-way ANOVA showed significant group differences in gender stigma consciousness, $F(2, 1863) = 41.00$, $p < 0.001$. Tukey post hoc tests indicated that the positive discrepancy group reported significantly higher gender stigma consciousness than both the zero and negative discrepancy groups (both $p < 0.001$), whereas the negative and zero discrepancy groups did not differ significantly. Together, these results suggest that students who perceived themselves as more masculine than they believed others perceived them reported lower belonging and heightened gender stigma consciousness, consistent with the correlational patterns shown in Table VI.

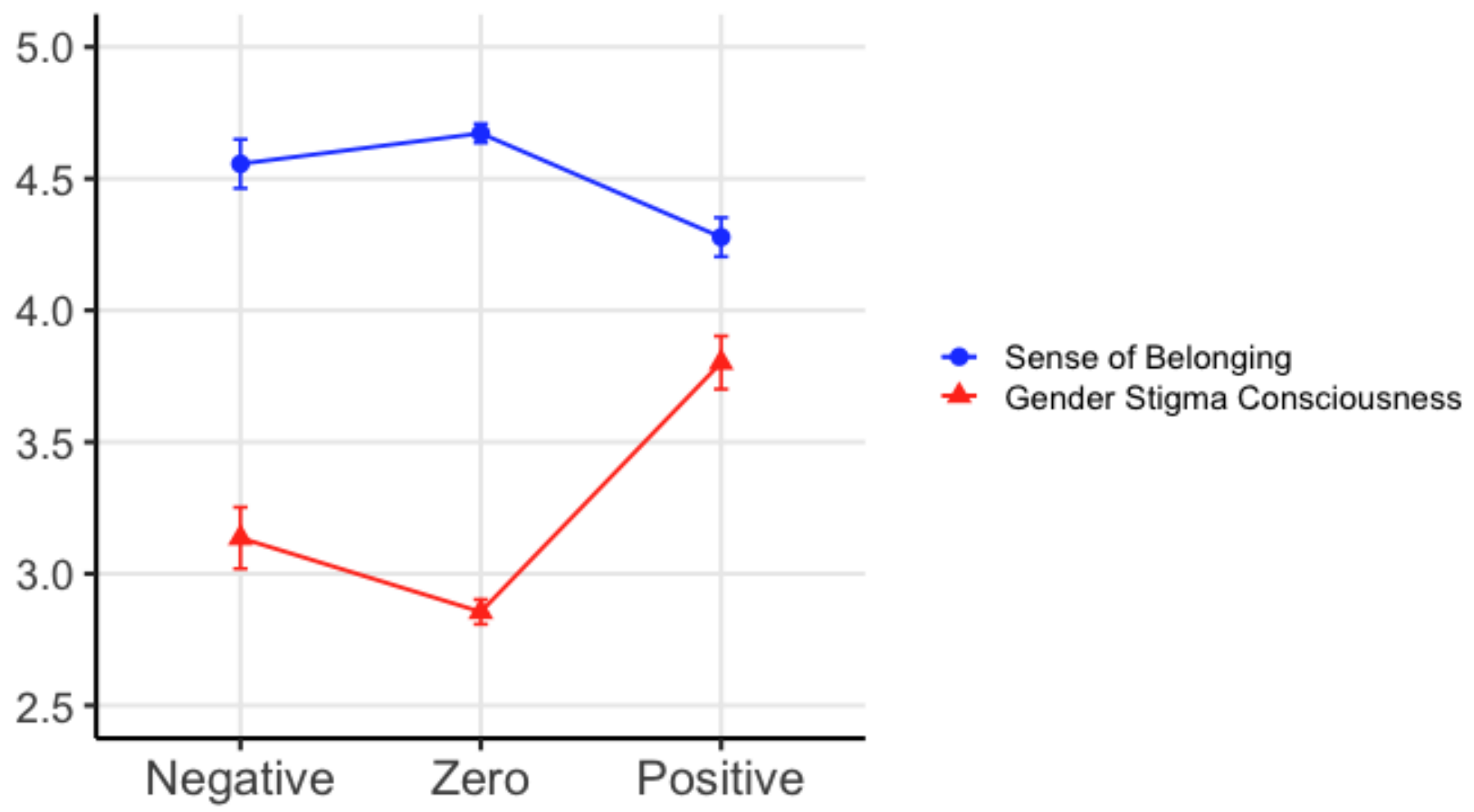

**Figure 4.** Mean sense of belonging and gender stigma consciousness across masculinity discrepancy groups. Error bars represent standard errors.

Figure 5 presents parallel comparisons of sense of belonging and gender stigma consciousness across the three femininity discrepancy groups. Students in the negative femininity discrepancy group—those whose self-appraised femininity was lower than their reflected appraisal—showed the lowest sense of belonging and the highest gender stigma consciousness across the three groups.

A one-way ANOVA indicated significant group differences in sense of belonging, $F(2, 1800) = 14.30$, $p < 0.001$. Tukey post hoc tests showed that both the zero and positive discrepancy groups reported significantly higher belonging than the negative discrepancy group (both $p < 0.001$), whereas the zero and positive groups did not differ significantly ($p=0.306$).

Gender stigma consciousness also differed significantly across femininity discrepancy groups, $F(2, 1808) = 30.20$, $p < 0.001$. Tukey post hoc tests showed that the negative discrepancy group reported significantly higher gender stigma consciousness than both the zero and positive discrepancy groups (both $p < 0.001$), whereas the zero and positive groups did not differ significantly ($p = 0.201$). Together, these results suggest that negative femininity discrepancy was associated with lower belonging and heightened gender stigma consciousness, consistent with the correlational patterns shown in Table VI.

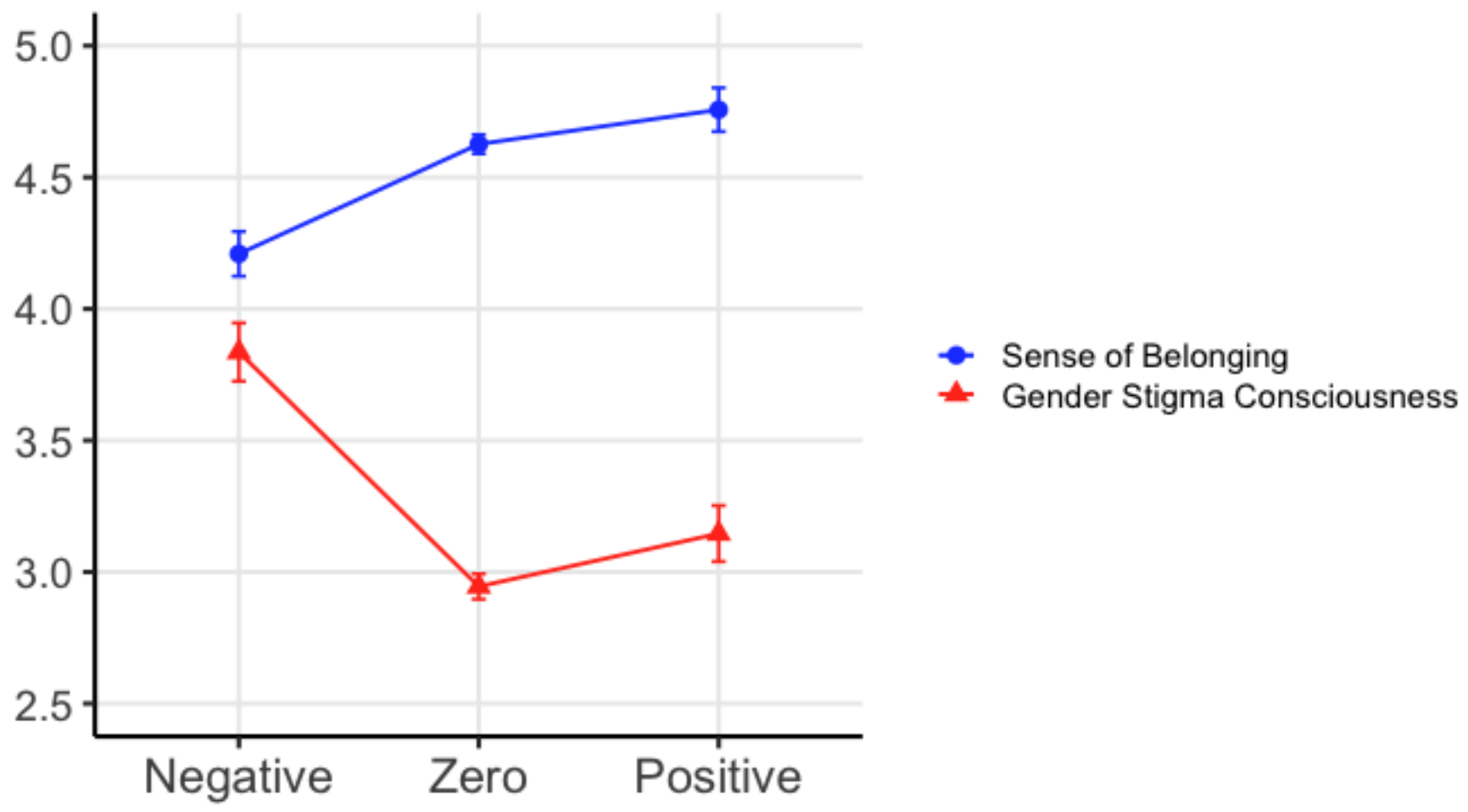

**Figure 5.** Mean sense of belonging and gender stigma consciousness across femininity discrepancy groups. Error bars represent standard errors.

Figure 6 presents parallel comparisons of sense of belonging and gender stigma consciousness across the three androgyny discrepancy groups. Unlike the masculinity and femininity results shown in Figures 4 and 5, sense of belonging was relatively similar across the three androgyny discrepancy groups. A one-way ANOVA indicated no significant group differences in sense of belonging, $F(2, 1676) = 1.11$, $p = 0.330$.

Gender stigma consciousness showed a different pattern, with significant differences across androgyny discrepancy groups, $F(2, 1684) = 52.30$, $p < 0.001$. Tukey post hoc tests showed that the positive discrepancy group reported significantly higher gender stigma consciousness than both the zero discrepancy group ($p < 0.001$) and the negative discrepancy group ($p = 0.008$). The negative discrepancy group also reported significantly higher gender stigma consciousness than the zero discrepancy group ($p = 0.004$). Together, these results suggest that androgyny discrepancies were not associated with differences in sense of belonging, but were associated with heightened gender stigma consciousness, consistent with the correlational patterns shown in Table VI.

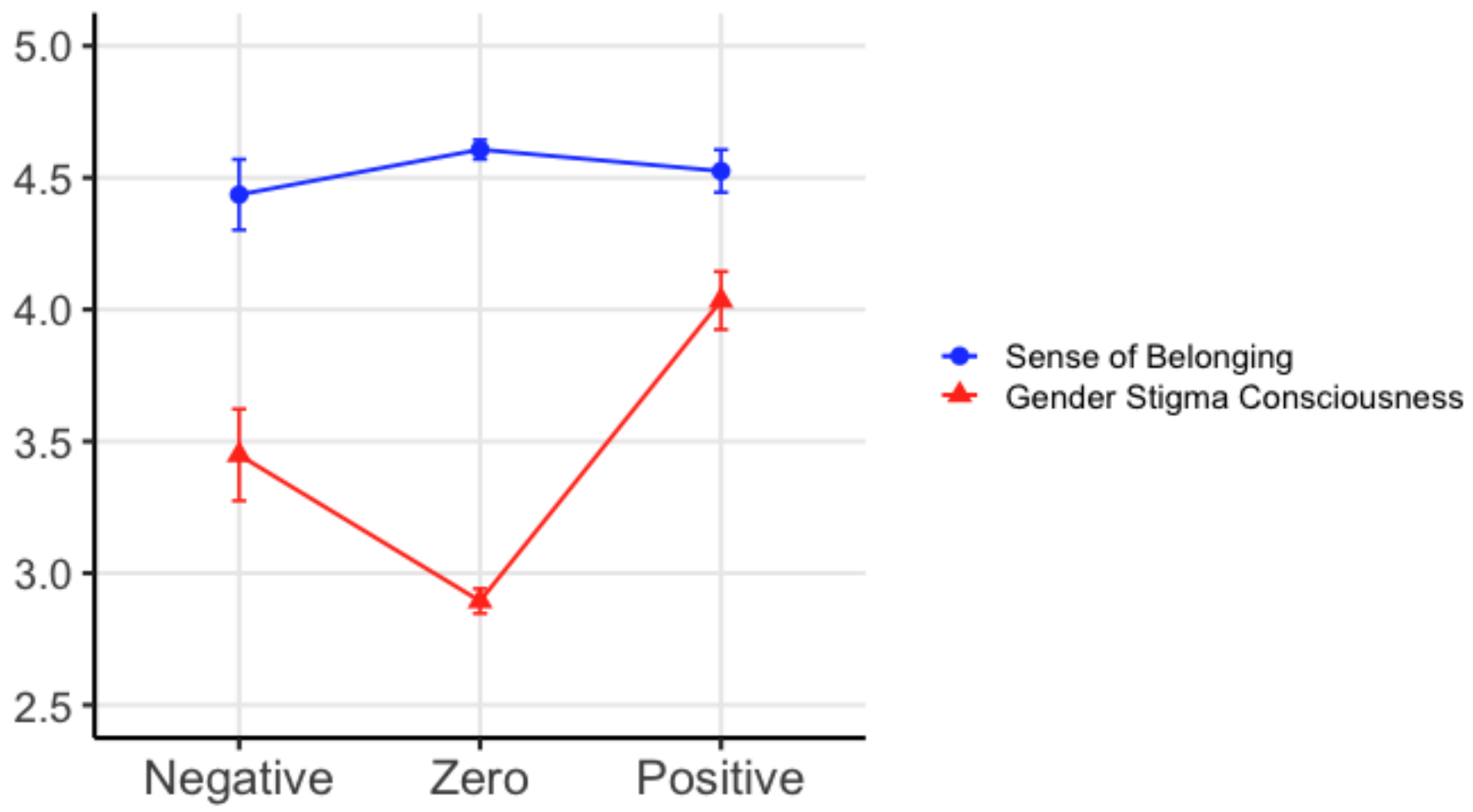

**Figure 6.** Mean sense of belonging and gender stigma consciousness across androgyny discrepancy groups. Error bars represent standard errors.

The preceding analyses showed that self–reflected appraisal discrepancies were associated with students' sense of belonging and gender stigma consciousness, and that these social-psychological variables were also correlated with students' demographic variables and course type. Before fitting the subsequent logistic regression models predicting appraisal discrepancy indicators, we examined baseline associations among these variables using two multiple linear regression models predicting gender stigma consciousness and sense of belonging from students' gender, race/ethnicity, and course type. Table VII presents the unstandardized regression coefficients for these models. In both models, coefficients represent associations after controlling for the other predictors in the model.

As shown in Table VII, the Gender variable predicted higher gender stigma consciousness and lower sense of belonging: women and students in gender identity groups with smaller sample sizes reported higher gender stigma consciousness ($b = 1.41$, $SE = 0.08$, $p < 0.001$) and lower sense of belonging ($b = -0.62$, $SE = 0.06$, $p < 0.001$) than men. Course type also predicted both outcomes, with students in calculus-based courses reporting higher gender stigma consciousness ($b = 0.50$, $SE = 0.09$, $p < 0.001$) and slightly higher sense of belonging ($b = 0.14$, $SE = 0.07$, $p < 0.05$). Race/ethnicity predicted sense of belonging but not gender stigma consciousness, with students from other racial/ethnic groups reporting lower belonging than students who identified as White ($b = -0.22$, $SE = 0.06$, $p < 0.001$).

**TABLE VII.** Unstandardized coefficients from two linear regression models predicting gender stigma consciousness and sense of belonging, respectively. Standard errors are shown in parentheses. Statistical significance is indicated by *** for p < 0.001, ** for 0.001 ≤ p < 0.01, * for 0.01 ≤ p < 0.05, and ns for p ≥ 0.05.

| | Predicted variable | |
|---|---|---|
| | Gender stigma consciousness | Sense of belonging |
| **Predictive variable** | | |
| Gender | 1.41 (0.08) *** | -0.62 (0.06) *** |
| Race | 0.12 (0.08) ns | -0.22 (0.06) *** |
| Course type | 0.50 (0.09) *** | 0.14 (0.07) * |

## D. Logistic regression models predicting self–reflected appraisal discrepancies in masculinity, femininity, and androgyny

The preceding analyses showed that self–reflected appraisal discrepancies were associated with demographic, course-type, and social-psychological variables. Because these variables were interrelated, we next used logistic regression models to examine how they jointly predicted the likelihood of reporting directional discrepancies in masculinity, femininity, and androgyny.

Table VIII presents a series of logistic regression models predicting the likelihood of reporting a positive masculinity discrepancy, defined as having a self-appraised masculinity score higher than the corresponding reflected appraisal. Model 1 includes demographic predictors, Gender and Race/ethnicity; Model 2 adds course type; and Model 3 further incorporates the social-psychological variables sense of belonging and gender stigma consciousness.

In Model 1, the Gender variable was a strong predictor of positive masculinity discrepancy (OR = 2.93, p < 0.001). Students who identified as women or were in gender identity groups with smaller sample sizes had 2.93 times the odds of reporting a positive masculinity discrepancy compared with students who identified as men. Race/ethnicity was not a statistically significant predictor. This pattern remained stable in Model 2 after adding course type: the odds ratio for Gender changed only slightly (OR = 3.00, p < 0.001), and course type was not statistically significant.

Model 3 builds on Model 2 by adding sense of belonging and gender stigma consciousness. Both social-psychological variables were statistically significant predictors after controlling for the other variables in the model. Higher gender stigma consciousness was associated with increased odds of reporting a positive masculinity discrepancy (OR = 1.24, p < 0.001), meaning that each one-unit increase in stigma consciousness corresponded to a 24% increase in the odds of reporting this discrepancy. In contrast, higher sense of belonging was associated with reduced odds of reporting a positive masculinity discrepancy (OR = 0.90, p = 0.026), indicating that each one-unit increase in belonging corresponded to a 10% decrease in the odds of reporting this discrepancy. In addition, the odds ratio for Gender decreased from 3.00 to 2.07 after sense of belonging and gender stigma consciousness were included, suggesting these social-psychological variables may account for part of the association between the Gender variable and positive masculinity discrepancy. Model fit also improved after sense of belonging and gender stigma consciousness were added, as indicated by the decrease in AIC [92] from Model 2 to Model 3. This suggests that including these social-psychological variables improved model fit beyond models with demographic and course-level factors alone.

**TABLE VIII.** Logistic regression models predicting positive masculinity discrepancy (i.e., self-appraisal of masculinity > reflected appraisal of masculinity). Odds ratios, p values, and Akaike information criterion (AIC) values are reported.

| Predictor | Model 1 | | Model 2 | | Model 3 | |
|---|---|---|---|---|---|---|
| | *p*-value | Odds Ratio | *p*-value | Odds Ratio | *p*-value | Odds Ratio |
| Intercept | <0.001 | 0.14 | <0.001 | 0.13 | <0.001 | 0.15 |
| Gender | <0.001 | 2.93 | <0.001 | 3.00 | <0.001 | 2.07 |
| Race/ethnicity | 0.241 | 0.86 | 0.240 | 0.86 | 0.163 | 0.83 |
| Course type | - | - | 0.650 | 1.07 | 0.714 | 0.95 |
| Sense of belonging | - | - | - | - | 0.026 | 0.90 |
| Gender stigma consciousness | - | - | - | - | <0.001 | 1.24 |
| AIC | 1670.53 | | 1672.32 | | 1620.48 | |

Table IX presents a series of logistic regression models predicting the likelihood of reporting a negative femininity discrepancy, defined as having a self-appraised femininity score lower than the corresponding reflected appraisal. Model 4 includes demographic predictors, Model 5 adds course type, and Model 6 further incorporates sense of belonging and gender stigma consciousness.

In Model 4, the Gender variable was a strong predictor of negative femininity discrepancy (OR = 2.80, $p < 0.001$). Students who identified as women or were in gender identity groups with smaller sample sizes had 2.80 times the odds of reporting this discrepancy compared with students who identified as men. Race/ethnicity was not a statistically significant predictor. After course type was added in Model 5, the Gender variable remained a strong predictor, whereas neither race/ethnicity nor course type was statistically significant.

Model 6 adds sense of belonging and gender stigma consciousness, both of which were statistically significant predictors. Higher gender stigma consciousness was associated with increased odds of reporting a negative femininity discrepancy (OR = 1.20, $p < 0.001$), whereas higher sense of belonging was associated with reduced odds (OR = 0.88, $p = 0.016$). The odds ratio for Gender decreased from 2.92 in Model 5 to 2.09 in Model 6, suggesting that part of the association between the Gender variable and femininity discrepancy may be accounted for by these social-psychological variables. Model fit also improved after the social-psychological variables were added, as reflected in the decrease in AIC from 1466.08 in Model 5 to 1426.55 in Model 6.

**TABLE IX.** Logistic regression models predicting negative femininity discrepancy (i.e., self-appraisal of femininity < reflected appraisal of femininity). Odds ratios, p values, and Akaike information criterion (AIC) values are reported

| Predictor | Model 4 | | Model 5 | | Model 6 | |
|---|---|---|---|---|---|---|
| | *p*-value | Odds Ratio | *p*-value | Odds Ratio | *p*-value | Odds Ratio |
| Intercept | <0.001 | 0.10 | <0.001 | 0.09 | <0.001 | 0.11 |
| Gender | <0.001 | 2.80 | <0.001 | 2.92 | <0.001 | 2.09 |
| Race/Ethnicity | 0.693 | 1.06 | 0.688 | 1.06 | 0.804 | 1.04 |
| Course type | - | - | 0.413 | 1.14 | 0.879 | 1.02 |
| Sense of belonging | - | - | - | - | 0.016 | 0.88 |
| Gender stigma consciousness | - | - | - | - | <0.001 | 1.20 |
| AIC | 1464.76 | | 1466.08 | | 1426.55 | |

Table X presents a series of logistic regression models predicting the likelihood of reporting a positive androgyny discrepancy, defined as having a self-appraised androgyny score higher than the corresponding reflected appraisal. Model 7 includes demographic predictors, Model 8 adds course type, and Model 9 further incorporates sense of belonging and gender stigma consciousness.

In Model 7, the Gender variable was a strong predictor of reporting a positive androgyny discrepancy (OR = 2.73, $p < 0.001$). Students who identified as women or were in gender identity groups with smaller sample sizes had 2.73 times the odds of reporting this discrepancy compared with students who identified as men. Unlike the models for masculinity and femininity, Race/ethnicity was also a statistically significant predictor (OR = 0.68, $p = 0.009$), with students from other racial/ethnic groups showing lower odds of reporting this discrepancy than White students.

Model 8 builds on Model 7 by adding course type. In contrast to the masculinity and femininity models, enrollment in calculus-based courses was associated with higher odds of reporting a positive androgyny discrepancy (OR = 1.52, $p = 0.011$). The Gender variable remained a strong predictor (OR = 3.10, $p < 0.001$), and Race/ethnicity also remained statistically significant (OR = 0.67, $p = 0.008$).

Model 9 adds sense of belonging and gender stigma consciousness. Gender stigma consciousness was a statistically significant predictor (OR = 1.33, $p < 0.001$), whereas sense of belonging was not statistically significant. In addition, the inclusion of social-psychological variables improved model fit, as reflected in the reduction in AIC from 1409.59 in Model 8 to 1337.57 in Model 9.

Across the three discrepancy outcomes, the Gender variable remained a consistent predictor, while gender stigma consciousness was the most consistent social-psychological predictor. Sense of belonging was a significant predictor of positive masculinity and negative femininity discrepancies, but not of positive androgyny discrepancy.

**TABLE X.** Logistic regression models predicting positive androgyny discrepancy (i.e., self-appraisal of androgyny > reflected appraisal of androgyny). Odds ratios, p values, and Akaike information criterion (AIC) values are reported

| Predictor | Model 7 | | Model 8 | | Model 9 | |
|---|---|---|---|---|---|---|
| | *p*-value | Odds Ratio | *p*-value | Odds Ratio | *p*-value | Odds Ratio |
| Intercept | <0.001 | 0.13 | <0.001 | 0.09 | <0.001 | 0.03 |
| Gender | <0.001 | 2.73 | <0.001 | 3.10 | <0.001 | 2.19 |
| Race/Ethnicity | 0.009 | 0.68 | 0.008 | 0.67 | 0.009 | 0.67 |
| Course type | - | - | 0.011 | 1.52 | 0.111 | 1.31 |
| Sense of belonging | - | - | - | - | 0.115 | 1.09 |
| Gender stigma consciousness | - | - | - | - | <0.001 | 1.33 |
| AIC | 1414.24 | | 1409.59 | | 1337.57 | |

### E. Cross-Institutional Comparison

In this section, we compare results from the present study with those reported in the prior quantitative study that the present work conceptually replicates [43]. The comparison focuses on variation in students' gradational ratings of femininity, masculinity, and androgyny, patterns of self–reflected appraisal discrepancies, and predictors of discrepancy outcomes across institutional contexts. Table XI summarizes the major similarities and differences.

In both studies, students used the full range of response options on the gradational scales, and substantial within-group variation was evident across femininity, masculinity, and androgyny. The overall patterns of femininity, masculinity, and androgyny across gender groups were also broadly similar across institutions. Compared with the prior study, responses in the present study were less concentrated at the endpoints and more concentrated at intermediate values, and the present study also included students in gender identity groups with smaller sample sizes.

At the level of self–reflected discrepancies, several common patterns emerged across the two studies. In both studies, women showed tendencies toward positive masculinity, negative femininity, and positive androgyny discrepancies. At the same time, some differences were observed in how these patterns appeared across gender groups. For example, positive masculinity discrepancy was observed only among women in the present study, whereas in the prior study it was observed among both women and men. For femininity, women showed negative discrepancy in both studies, but in the present study this pattern was accompanied by positive discrepancy among men and negative discrepancy among students in smaller-sample gender identity groups. For androgyny, the positive discrepancy observed among women in the prior study was also observed in the present study, and positive androgyny discrepancy was also found among students in smaller-sample gender identity groups and, more weakly, among men.

The logistic regression analyses also showed several common patterns across institutions. In both studies, the gender variable and gender stigma consciousness were significant predictors of positive masculinity discrepancy. In both studies, the gender variable, lower sense of belonging, and higher gender stigma consciousness were significant predictors of negative femininity discrepancy. At the same time, some differences were also observed. In the present study, the gender variable showed stronger associations with the discrepancy outcomes, as reflected in larger odds ratios, than in the prior study. Lower sense of belonging also emerged as a significant predictor of positive masculinity discrepancy in the present study. For positive androgyny discrepancy, the roles of race/ethnicity and course type differed across the two studies. In the prior study, course type was a significant predictor, whereas in the present study it was no longer statistically significant after the social-psychological variables were included. In addition, race/ethnicity was significantly associated with positive androgyny discrepancy in both studies, but in opposite directions: students in the combined racial/ethnic minority group showed higher odds of positive androgyny discrepancy in the prior study but lower odds in the present study.

**TABLE XI.** Cross-institutional comparison of gender expression appraisal ratings, self–reflected appraisal discrepancies, and predictors of discrepancy outcomes in the prior and present studies.

| Comparison domain | Shared pattern across studies | Key differences between studies |
|---|---|---|
| **Overall variation in gender expression appraisal ratings** | In both studies, students used all response options on the 7-point gradational scales for femininity, masculinity, and androgyny, indicating substantial within-group variation. | Compared with the prior study, responses in the present study were less concentrated at the endpoints of the gradational scales and more concentrated at intermediate values. In addition, the present study included students in smaller-sample gender identity groups, whose responses also spanned the full range of response options. |
| **Masculinity discrepancy pattern** | In both studies, women showed a statistically significant positive masculinity discrepancy. | In the prior study, positive masculinity discrepancies were observed among both women and men. In the present study, such discrepancies were observed only among women, with women showing positive discrepancies in both course types while men and smaller-sample gender identity groups showed no statistically significant discrepancy pattern. |
| **Femininity discrepancy pattern** | In both studies, women showed a statistically significant negative femininity discrepancy. | In the prior study, negative femininity discrepancy was observed among women in calculus-based courses. In the present study, it was observed among women in algebra-based courses and among smaller-sample gender identity groups in calculus-based courses. |
| **Androgyny discrepancy pattern** | In both studies, women showed a statistically significant positive androgyny discrepancy. | In the prior study, positive androgyny discrepancy was observed only among women. In the present study, this pattern was observed among women and smaller-sample gender identity groups in both course types, and among men in calculus-based courses. |
| **Predictors of masculinity discrepancy** | In both studies, the gender variable and higher gender stigma consciousness were statistically significant predictors of positive masculinity discrepancy. | In the present study, the gender variable was a stronger predictor than in the prior study, and lower sense of belonging also emerged as a statistically significant predictor of positive masculinity discrepancy. |
| **Predictors of femininity discrepancy** | In both studies, the gender variable, lower sense of belonging, and higher gender stigma consciousness were statistically significant predictors of negative femininity discrepancy. | In the present study, the same overall predictor structure was observed, but the gender variable was a stronger predictor than in the prior study. |
| **Predictors of androgyny discrepancy** | In both studies, the gender variable and higher gender stigma consciousness were statistically significant predictors of positive androgyny discrepancy. | In the present study, the gender variable and gender stigma consciousness were stronger predictors than in the prior study. The roles of race/ethnicity and course type also differed across the two studies in direction or strength. |

## VII. Discussion

This study examined whether patterns identified in a prior quantitative study of students' self-appraisals and reflected appraisals of femininity, masculinity, and androgyny recur in a second institutional context [43]. Building on the cross-institutional comparison reported above, the present findings suggest both continuity and variation across institutional contexts. Several broad patterns recurred across the two studies, including substantial within-gender variation in students' appraisals, directional self–reflected appraisal discrepancies, and associations between discrepancy patterns and gender stigma consciousness. These recurring patterns suggest that several aspects of the prior findings are not limited to a single institutional dataset. At the same time, the two studies also differed in several ways, including the prevalence of endpoint versus intermediate responses and the roles of race/ethnicity, course type, and sense of belonging in some models. This variation is consistent with the view that self-appraisals and reflected appraisals along these gendered dimensions are socially situated [32,33]. Although the present data do not allow us to determine which specific contextual features underlie the observed differences, the comparison suggests that students' gendered self-appraisals and reflected appraisals may reflect both recurring features of physics learning environments and local contextual conditions. In the sections that follow, we discuss self–reflected appraisal discrepancy as an inclusion-relevant construct, possible interpretations of the observed discrepancy patterns, implications for physics education research and learning environments, and limitations and directions for future work.

### A. Self–reflected appraisal discrepancy and students' experiences of inclusion

Across the two institutional contexts, self–reflected appraisal discrepancies showed directional patterns that varied across gender groups. Several patterns recurred across institutions, especially among women, for whom positive masculinity discrepancy, negative femininity discrepancy, and positive androgyny discrepancy appeared across the two studies. These discrepancy patterns were also associated with students' social-psychological experiences in physics. Taken together, the gendered directionality of these discrepancies and their associations with belonging and gender stigma consciousness suggest that self–reflected appraisal discrepancy may provide a useful lens for examining aspects of students' experiences that are relevant to inclusion in physics.

This interpretation is informed by prior literature on sense of belonging and gender stigma consciousness. In physics education research, sense of belonging has been shown to predict students' academic performance [18,72], self-efficacy [47], physics identity [71], and persistence-related outcomes [48,73]. Prior research has linked gender stigma consciousness to impostor phenomenon [83], academic disengagement [93], lower academic self-concept [83], and stereotype threat [74]. Because sense of belonging and gender stigma consciousness are closely tied to students' experiences of fit, legitimacy, and inclusion, their associations with discrepancy patterns suggest that self–reflected appraisal discrepancy may be more than a descriptive mismatch between self-appraisals and reflected appraisals. Rather, such discrepancies may be related to students' concerns about whether aspects of how they see themselves are recognized as fitting within physics learning environments. The recurrence of these patterns across institutional settings further suggests that this phenomenon is not confined to a single context.

### B. Tentative interpretations of appraisal discrepancies

One possible interpretation is that these discrepancies may reflect how students navigate gendered expectations about competence, recognition, and fit in physics learning environments. This interpretation is motivated by the present and prior findings that these directional discrepancies were not evenly distributed across gender groups and appeared more frequently among women. Prior work has shown that physics culture is often associated with masculinity and that competence in physics is frequently linked to stereotypically masculine traits [35,37,94]. Prior qualitative research further suggests that women and gender minority students may experience pressure to manage how they present themselves in physics spaces, including distancing themselves from femininity [35,37], negotiating masculine-coded norms of competence [37], or navigating the risk that their gender expression will be misread or treated in exclusionary ways [41].

From this perspective, positive masculinity discrepancies may reflect a perceived gap in recognition. In a physics culture where competence and fit are often associated with masculinity, such discrepancies may indicate that students perceive aspects of themselves that could support a sense of fit in physics as not fully recognized by others. Negative femininity discrepancy may reflect a complementary concern. Some students may feel that their femininity is more visible or more readily emphasized by others than it is in their own self-understanding. Such perceived emphasis may shape whether they feel recognized as competent or legitimate participants in physics. Together, these patterns may point to a tension between how students understand themselves and how they believe they are positioned by others in relation to gendered expectations of physics. These interpretations are necessarily tentative because the present survey data do not show why students reported these discrepancies, whether they experienced them as problematic, or how specific classroom interactions shaped them. Future qualitative work is needed to examine how students interpret these mismatches in their own words and how they connect them to recognition, gender stereotypes, and belonging in physics.

### C. Implications for theory, PER, and learning environments

The present findings suggest that students' self-appraisals and reflected appraisals of femininity, masculinity, and androgyny may serve as a complement to categorical gender identity measures in studies of gender-related recognition and inclusion in physics. Whereas categorical gender identity measures are important for documenting group-level inequities, gradational self-appraisal and reflected-appraisal measures can reveal variation within gender groups and self–reflected appraisal discrepancies that might otherwise remain less visible [32,43]. These findings are broadly consistent with gender performativity theory [57], which treats gender as socially expressed, interpreted, and negotiated within specific contexts. From this perspective, self–reflected appraisal discrepancy may provide an analytic lens for examining perceived mismatches between students' self-appraisals and how they believe they are recognized in physics learning environments. Future research could examine whether different instructional contexts, classroom norms, and pedagogical approaches are associated with these discrepancies, and how these discrepancies relate to students' participation, motivation, academic outcomes, and persistence in physics.

These findings also have implications for the design of physics learning environments. If self–reflected appraisal discrepancy is related to concerns about recognition and social fit, then inclusive learning environments may need to support students not only in expressing and participating in ways that feel authentic to them, but also in having those forms of expression and participation recognized as legitimate ways of belonging and demonstrating competence in physics. This may include creating classroom norms that broaden what counts as competent participation, so that students can be recognized through multiple forms of engagement rather than primarily through speed, confidence, or public correctness. Such norms could, for example, value reasoning, questioning, evidence use, collaborative sensemaking, and responsiveness to peers' ideas [95,96], while reducing the extent to which competence is tied to narrow or gendered expectations of participation [35,37]. In this way, inclusive learning environments may support students' sense that their ways of participating are valued and legitimate within physics.

### D. Limitations and future directions

The present findings also point to several important directions for future research. Although the observed associations among appraisal discrepancy, sense of belonging, and gender stigma consciousness are informative, the present study does not establish their causal or temporal ordering. It remains unclear whether appraisal discrepancy contributes to lower belonging and heightened gender stigma consciousness, whether lower belonging and heightened gender stigma consciousness instead heighten students' perceptions of discrepancy, or whether all three are shaped by broader features of physics learning environments. In this sense, discrepancy may function as a mechanism, an outcome, or both. Clarifying these possibilities will require future work, particularly qualitative and longitudinal studies that examine how students interpret and experience these discrepancies in practice.

A second limitation concerns the small sample sizes for some gender identity categories. Although the present study included students who identified as non-binary, agender, or self-described their gender, these groups were too small to support separate statistical analyses. As a result, some analyses combined students from these gender identity categories with women for statistical stability; this analytic decision should not be interpreted as suggesting that these identities or experiences are interchangeable. Future qualitative and mixed-methods work is needed to examine how students with specific gender identities experience self-appraisals, reflected appraisals, belonging, and inclusion in physics.

Finally, although the present study extends prior work beyond a single institution, both institutions examined are large public research universities. Future research should examine whether similar patterns emerge in settings that differ in institutional type, student population, curricular structure, and regional context.

References

[1] *Digest of Education Statistics, 2013*, https://nces.ed.gov/programs/digest/d13/tables/dt13_318.10.asp.

[2] A. M. Porter, *How Women Persist in Undergraduate Physics*, https://www.aip.org/statistics/how-women-persist-in-undergraduate-physics.

[3] A. Maries, N. I. Karim, and C. Singh, Active Learning in an Inequitable Learning Environment Can Increase the Gender Performance Gap: The Negative Impact of Stereotype Threat, Phys. Teach. **58**, 430 (2020).

[4] M. Dew, J. Perry, L. Ford, W. Bassichis, and T. Erukhimova, Gendered performance differences in introductory physics: A study from a large land-grant university, Phys. Rev. Phys. Educ. Res. **17**, 010106 (2021).

[5] L. McCullough, Gender, context, and physics assessment, J. Int. Womens Stud. **5**, (2004).

[6] M. Lorenzo, C. H. Crouch, and E. Mazur, Reducing the gender gap in the physics classroom, Am. J. Phys. **74**, 118 (2006).

[7] S. J. Pollock, N. D. Finkelstein, and L. E. Kost, Reducing the gender gap in the physics classroom: How sufficient is interactive engagement?, Phys. Rev. Spec. Top. - Phys. Educ. Res. **3**, 010107 (2007).

[8] L. E. Kost, S. J. Pollock, and N. D. Finkelstein, Characterizing the gender gap in introductory physics, Phys. Rev. Spec. Top. - Phys. Educ. Res. **5**, 010101 (2009).

[9] L. E. Kost-Smith, S. J. Pollock, and N. D. Finkelstein, Gender disparities in second-semester college physics: The incremental effects of a ``smog of bias'', Phys. Rev. Spec. Top. - Phys. Educ. Res. **6**, 020112 (2010).

[10] A. Madsen, S. B. McKagan, and E. C. Sayre, Gender gap on concept inventories in physics: What is consistent, what is inconsistent, and what factors influence the gap?, Phys. Rev. Spec. Top. - Phys. Educ. Res. **9**, 020121 (2013).

[11] R. Henderson, G. Stewart, J. Stewart, L. Michaluk, and A. Traxler, Exploring the gender gap in the conceptual survey of electricity and magnetism, Phys. Rev. Phys. Educ. Res. **13**, 020114 (2017).

[12] N. I. Karim, A. Maries, and C. Singh, Do evidence-based active-engagement courses reduce the gender gap in introductory physics?, Eur. J. Phys. **39**, 025701 (2018).

[13] R. Henderson, P. Miller, J. Stewart, A. Traxler, and R. Lindell, Item-level gender fairness in the Force and Motion Conceptual Evaluation and the Conceptual Survey of Electricity and Magnetism, Phys. Rev. Phys. Educ. Res. **14**, 020103 (2018).

[14] A. Traxler, R. Henderson, J. Stewart, G. Stewart, A. Papak, and R. Lindell, Gender fairness within the Force Concept Inventory, Phys. Rev. Phys. Educ. Res. **14**, 010103 (2018).

[15] M. Mears, Gender differences in the Force Concept Inventory for different educational levels in the United Kingdom, Phys. Rev. Phys. Educ. Res. **15**, 020135 (2019).

[16] A. Miyake, L. E. Kost-Smith, N. D. Finkelstein, S. J. Pollock, G. L. Cohen, and T. A. Ito, Reducing the Gender Achievement Gap in College Science: A Classroom Study of Values Affirmation, Science **330**, 1234 (2010).

[17] R. L. Matz, B. P. Koester, S. Fiorini, G. Grom, L. Shepard, C. G. Stangor, B. Weiner, and T. A. McKay, Patterns of Gendered Performance Differences in Large Introductory Courses at Five Research Universities, AERA Open **3**, 2332858417743754 (2017).

[18] Y. Li and C. Singh, Sense of belonging is an important predictor of introductory physics students' academic performance, Phys. Rev. Phys. Educ. Res. **19**, 020137 (2023).

[19] S. Salehi, E. Burkholder, G. P. Lepage, S. Pollock, and C. Wieman, Demographic gaps or preparation gaps?: The large impact of incoming preparation on performance of students in introductory physics, Phys. Rev. Phys. Educ. Res. **15**, 020114 (2019).
[20] C. Lindstrøm and M. D. Sharma, Self-Efficacy of First Year University Physics Students: Do Gender and Prior Formal Instruction in Physics Matter?, Int. J. Innov. Sci. Math. Educ. **19**, (2011).
[21] R. S. Barthelemy and A. V. Knaub, Gendered motivations and aspirations of university physics students in Finland, Phys. Rev. Phys. Educ. Res. **16**, 010133 (2020).
[22] J. M. Nissen and J. T. Shemwell, Gender, experience, and self-efficacy in introductory physics, Phys. Rev. Phys. Educ. Res. **12**, 020105 (2016).
[23] Y. Li and C. Singh, Do female and male students' physics motivational beliefs change in a two-semester introductory physics course sequence?, Phys. Rev. Phys. Educ. Res. **18**, 010142 (2022).
[24] Z. Hazari, G. Sonnert, P. M. Sadler, and M. Shanahan, Connecting high school physics experiences, outcome expectations, physics identity, and physics career choice: A gender study, J. Res. Sci. Teach. **47**, 978 (2010).
[25] K. L. Lewis, J. G. Stout, S. J. Pollock, N. D. Finkelstein, and T. A. Ito, Fitting in or opting out: A review of key social-psychological factors influencing a sense of belonging for women in physics, Phys. Rev. Phys. Educ. Res. **12**, 020110 (2016).
[26] A. Bandura, *Self-Efficacy*, in *Encyclopedia of Psychology, Vol. 7.* (American Psychological Association, Washington, DC, US, 2000), pp. 212–213.
[27] R. Ivie, S. White, and R. Y. Chu, Women's and men's career choices in astronomy and astrophysics, Phys. Rev. Phys. Educ. Res. **12**, 020109 (2016).
[28] V. Sawtelle, E. Brewe, and L. H. Kramer, Exploring the relationship between self-efficacy and retention in introductory physics, J. Res. Sci. Teach. **49**, 1096 (2012).
[29] L. J. Sax, K. J. Lehman, R. S. Barthelemy, and G. Lim, Women in physics: A comparison to science, technology, engineering, and math education over four decades, Phys. Rev. Phys. Educ. Res. **12**, 020108 (2016).
[30] C. G. Hart, A. Saperstein, D. Magliozzi, and L. Westbrook, Gender and Health: Beyond Binary Categorical Measurement, J. Health Soc. Behav. **60**, 101 (2019).
[31] A. L. Traxler, X. C. Cid, J. Blue, and R. Barthelemy, Enriching gender in physics education research: A binary past and a complex future, Phys. Rev. Phys. Educ. Res. **12**, 020114 (2016).
[32] D. Magliozzi, A. Saperstein, and L. Westbrook, Scaling Up: Representing Gender Diversity in Survey Research, Socius **2**, 2378023116664352 (2016).
[33] L. Westbrook and K. Schilt, Doing Gender, Determining Gender: Transgender People, Gender Panics, and the Maintenance of the Sex/Gender/Sexuality System, Gend. Soc. **28**, 32 (2014).
[34] J. Butler, *Gender Trouble: Tenth Anniversary Edition*, 2nd ed. (Routledge, New York, 2002).
[35] A. T. Danielsson, Exploring woman university physics students 'doing gender' and 'doing physics,' Gend. Educ. **24**, 25 (2012).
[36] S. J. Kessler and W. McKenna, *Gender: An Ethnomethodological Approach* (University of Chicago Press, 1985).

[37] A. J. Gonsalves, “Physics and the girly girl—there is a contradiction somewhere”: doctoral students’ positioning around discourses of gender and competence in physics, Cult. Stud. Sci. Educ. **9**, 503 (2014).
[38] A. J. Gonsalves, A. Danielsson, and H. Pettersson, Masculinities and experimental practices in physics: The view from three case studies, Phys. Rev. Phys. Educ. Res. **12**, 020120 (2016).
[39] J. Maloy, M. B. Kwapisz, and B. E. Hughes, Factors Influencing Retention of Transgender and Gender Nonconforming Students in Undergraduate STEM Majors, CBE—Life Sci. Educ. **21**, ar13 (2022).
[40] R. S. Barthelemy, LGBT+ physicists qualitative experiences of exclusionary behavior and harassment, Eur. J. Phys. **41**, 065703 (2020).
[41] R. S. Barthelemy, M. Swirtz, S. Garmon, E. H. Simmons, K. Reeves, M. L. Falk, W. Deconinck, E. A. Long, and T. J. Atherton, $\mathrm{LGBT}+$ physicists: Harassment, persistence, and uneven support, Phys. Rev. Phys. Educ. Res. **18**, 010124 (2022).
[42] R. S. Barthelemy, B. E. Hughes, M. Swirtz, M. Mikota, and T. J. Atherton, Workplace climate for $\mathrm{LGBT}+$ physicists: A view from students and professional physicists, Phys. Rev. Phys. Educ. Res. **18**, 010147 (2022).
[43] Y. Li and E. Burkholder, Investigating students’ self-identified and reflected appraisal of femininity, masculinity, and androgyny in introductory physics courses, Phys. Rev. Phys. Educ. Res. **20**, 010110 (2024).
[44] J. A. Taylor and L. V. Hedges, *Toward More Rapid Accumulation of Knowledge about What Works in Physics Education: The Role of Replication, Reporting Practices, and Meta-Analysis*, in *The International Handbook of Physics Education Research: Special Topics*, edited by D. E. Meltzer, M. F. Taşar, and P. R. L. Heron (AIP Publishing LLC, 2023), p. 0.
[45] A. Maries, Y. Li, and C. Singh, Challenges faced by women and persons excluded because of their ethnicity and race in physics learning environments: review of the literature and recommendations for departments and instructors, Rep. Prog. Phys. **88**, 015901 (2025).
[46] R. Henderson, J. Stewart, and A. Traxler, Partitioning the gender gap in physics conceptual inventories: Force Concept Inventory, Force and Motion Conceptual Evaluation, and Conceptual Survey of Electricity and Magnetism, Phys. Rev. Phys. Educ. Res. **15**, 010131 (2019).
[47] Y. Li and C. Singh, Effect of gender, self-efficacy, and interest on perception of the learning environment and outcomes in calculus-based introductory physics courses, Phys. Rev. Phys. Educ. Res. **17**, 010143 (2021).
[48] K. L. Lewis, J. G. Stout, N. D. Finkelstein, S. J. Pollock, A. Miyake, G. L. Cohen, and T. A. Ito, Fitting in to Move Forward: Belonging, Gender, and Persistence in the Physical Sciences, Technology, Engineering, and Mathematics (pSTEM), Psychol. Women Q. **41**, 420 (2017).
[49] Y. Li and C. Singh, Impact of perceived recognition by physics instructors on women’s self-efficacy and interest, Phys. Rev. Phys. Educ. Res. **19**, 020125 (2023).
[50] J. G. Stout, N. Dasgupta, M. Hunsinger, and M. A. McManus, STEMing the tide: Using ingroup experts to inoculate women’s self-concept in science, technology, engineering, and mathematics (STEM), J. Pers. Soc. Psychol. **100**, 255 (2011).
[51] C. Frayling, *Mad, Bad and Dangerous?: The Scientist and the Cinema* (Reaktion Books, 2013).

[52] M. A. Weitekamp, *The Image of Scientists in The Big Bang Theory*, https://doi.org/10.1063/PT.3.3427.
[53] S. Upson and L. F. Friedman, Where Are All the Female Geniuses?, Sci. Am. Mind **23**, 63 (2012).
[54] L. Bian, S.-J. Leslie, and A. Cimpian, Gender stereotypes about intellectual ability emerge early and influence children's interests, Science **355**, 389 (2017).
[55] J. G. Stout and H. M. Wright, Lesbian, Gay, Bisexual, Transgender, and Queer Students' Sense of Belonging in Computing: An Intersectional Approach, Comput. Sci. Eng. **18**, 24 (2016).
[56] C. D. X. Hall, C. V. Wood, M. Hurtado, D. A. Moskowitz, C. Dyar, and B. Mustanski, Identifying leaks in the STEM recruitment pipeline among sexual and gender minority US secondary students, PLOS ONE **17**, e0268769 (2022).
[57] J. Butler, Feminism and the Subversion of Identity, N. Y. (1999).
[58] S. M. Anderson, Gender Matters: The Perceived Role of Gender Expression in Discrimination Against Cisgender and Transgender LGBQ Individuals, Psychol. Women Q. **44**, 323 (2020).
[59] A. M. Beltz, A. M. Loviska, and A. Weigard, Daily gender expression is associated with psychological adjustment for some people, but mainly men, Sci. Rep. **11**, 9114 (2021).
[60] L. M. Diamond, Gender Fluidity and Nonbinary Gender Identities Among Children and Adolescents, Child Dev. Perspect. **14**, 110 (2020).
[61] X. R. Quichocho, E. M. Schipull, and E. W. Close, *Understanding Physics Identity Development through the Identity Performances of Black, Indigenous, and Women of Color and LGBTQ+ Women in Physics*, in *2020 Physics Education Research Conference Proceedings* (American Association of Physics Teachers, Virtual Conference, 2020), pp. 412–417.
[62] W. Wood and A. H. Eagly, Two Traditions of Research on Gender Identity, Sex Roles **73**, 461 (2015).
[63] A. Constantinople, Masculinity-femininity: An exception to a famous dictum?, Psychol. Bull. **80**, 389 (1973).
[64] S. L. Bem, The measurement of psychological androgyny, J. Consult. Clin. Psychol. **42**, 155 (1974).
[65] M. Shang, The Cultural Adjustment of Chinese-Born Males to the American Masculinity Paradigm: Issues of Identity and Self-Image in First-Generation Chinese-American Male Immigrants - ProQuest, The Chicago School of Professional Psychology, 2016.
[66] E. Markstedt, L. Wängnerud, M. Solevid, and M. Djerf-Pierre, The subjective meaning of gender: how survey designs affect perceptions of femininity and masculinity, Eur. J. Polit. Gend. **4**, 51 (2021).
[67] K. J. Dockendorff and C. Geist, *Gender Trouble beyond the LGB and T: Gender Image and Experiences of Marginalization on Campus*, kg6f2_v1.
[68] K. Dockendorff and C. Geist, Beyond the Binary: Gender Image and Experiences of Marginalization on Campus, J. Crit. Scholarsh. High. Educ. Stud. Aff. **6**, (2022).
[69] D. Cassino and Y. Besen-Cassino, Political identity, gender identity or both? The political effects of sexual orientation and gender identity items in survey research, Eur. J. Polit. Gend. **4**, 71 (2021).
[70] A. Brenøe, Z. Eyibak, L. Heursen, E. Ranehill, and R. A. Weber, *Gender Identity and Economic Decision Making*, 5069992.

[71] Z. Hazari, D. Chari, G. Potvin, and E. Brewe, The context dependence of physics identity: Examining the role of performance/competence, recognition, interest, and sense of belonging for lower and upper female physics undergraduates, J. Res. Sci. Teach. **57**, 1583 (2020).
[72] S. Cwik and C. Singh, Students' sense of belonging in introductory physics course for bioscience majors predicts their grade, Phys. Rev. Phys. Educ. Res. **18**, 010139 (2022).
[73] J. G. Stout, T. A. Ito, N. D. Finkelstein, and S. J. Pollock, How a gender gap in belonging contributes to the gender gap in physics participation, AIP Conf. Proc. **1513**, 402 (2013).
[74] R. P. Brown and E. C. Pinel, Stigma on my mind: Individual differences in the experience of stereotype threat, J. Exp. Soc. Psychol. **39**, 626 (2003).
[75] J. Butler, Performative Acts and Gender Constitution: An Essay in Phenomenology and Feminist Theory, Theatre J. **40**, 519 (1988).
[76] J. A. Taylor and L. V. Hedges, Toward More Rapid Accumulation of Knowledge about What Works in Physics Education: The Role of Replication, Reporting Practices, and Meta-Analysis, (2023).
[77] M. C. Makel and J. A. Plucker, Facts Are More Important Than Novelty: Replication in the Education Sciences, Educ. Res. **43**, 304 (2014).
[78] B. Brandth, Agricultural body-building: Incorporations of gender, body and work, J. Rural Stud. **22**, 17 (2006).
[79] J. D. Smyth, A. Swendener, and E. Kazyak, Women's Work? The Relationship between Farmwork and Gender Self-Perception, Rural Sociol. **83**, 654 (2018).
[80] J. D. Smyth and K. Olson, *Male/Female Is Not Enough: Adding Measures of Masculinity and Femininity to General Population Surveys*, in *Understanding Survey Methodology: Sociological Theory and Applications*, edited by P. S. Brenner (Springer International Publishing, Cham, 2020), pp. 247–275.
[81] T. Morgenroth and M. K. Ryan, Gender Trouble in Social Psychology: How Can Butler's Work Inform Experimental Social Psychologists' Conceptualization of Gender?, Front. Psychol. **9**, (2018).
[82] L. R. Miller and E. A. Grollman, The Social Costs of Gender Nonconformity for Transgender Adults: Implications for Discrimination and Health, Sociol. Forum Randolph NJ **30**, 809 (2015).
[83] K. Cokley, G. Awad, L. Smith, S. Jackson, O. Awosogba, A. Hurst, S. Stone, L. Blondeau, and D. Roberts, The Roles of Gender Stigma Consciousness, Impostor Phenomenon and Academic Self-Concept in the Academic Outcomes of Women and Men, Sex Roles **73**, 414 (2015).
[84] E. Marshman, Z. Y. Kalender, C. Schunn, T. Nokes-Malach, and C. Singh, A longitudinal analysis of students' motivational characteristics in introductory physics courses: Gender differences, Can. J. Phys. **96**, 391 (2018).
[85] S. M. Aguillon, G.-F. Siegmund, R. H. Petipas, A. G. Drake, S. Cotner, and C. J. Ballen, Gender Differences in Student Participation in an Active-Learning Classroom, CBE—Life Sci. Educ. **19**, ar12 (2020).
[86] *Advance Student Success | Ascend | PERTS*, https://www.perts.net/ascend.
[87] J. Nunnally and I. Bernstein, *Psychometric Theory*, 3rd ed. (McGraw-Hill, New York, 1994).
[88] J. Cohen, *Statistical Power Analysis for the Behavioral Sciences*, 2nd ed. (Routledge, New York, NY, 1988).

[89] F. Wilcoxon, Individual Comparisons by Ranking Methods, Biom. Bull. **1**, 80 (1945).
[90] A. Agresti and B. Finlay, *Statistical Methods for the Social Sciences Pearson New International Edition* (Pearson International Edition, Harlow, Essex, 2009).
[91] A. M. Porter and R. Ivie, *Women in Physics and Astronomy, 2019*, https://www.aip.org/statistics/women-in-physics-and-astronomy-2019.
[92] H. Akaike, A new look at the statistical model identification, IEEE Trans. Autom. Control **19**, 716 (1974).
[93] E. C. Pinel, L. R. Warner, and P. Chua, Getting There is Only Half the Battle: Stigma Consciousness and Maintaining Diversity in Higher Education, J. Soc. Issues **61**, 481 (2005).
[94] S.-J. Leslie, A. Cimpian, M. Meyer, and E. Freeland, Expectations of brilliance underlie gender distributions across academic disciplines, Science **347**, 262 (2015).
[95] T. O. B. Odden and R. S. Russ, Defining sensemaking: Bringing clarity to a fragmented theoretical construct, Sci. Educ. **103**, 187 (2019).
[96] R. A. Engle and F. R. Conant, Guiding Principles for Fostering Productive Disciplinary Engagement: Explaining an Emergent Argument in a Community of Learners Classroom, Cogn. Instr. **20**, 399 (2002).